\documentclass{emulateapj}

\catcode`\"=\active\let"=\"

\def\3{{\ss} }

\def\c12{{1\over 2}}

\def\plusplus{\raise 0.3ex\hbox{${\scriptstyle ++}$}{}}

\newcommand{\oversim}[2]{\protect{\mbox{\lower0.5ex\vbox{%
   \baselineskip=0pt\lineskip=0.2ex
   \ialign{$\mathsurround=0pt #1\hfil##\hfil$\crcr#2\crcr\sim\crcr}}}}} 
\begin{document}

\title{Modeling Tidal Streams in evolving dark matter halos}
\author{Jorge Pe\~{n}arrubia$^1$
, Andrew J. Benson$^2$, David Mart\'inez-Delgado$^3$ \& Hans Walter Rix$^1$}
\affil{$^1$Max Planck Institut f\"ur Astronomie, K\"onigstuhl 17, Heidelberg, D-69117, Germany}
\affil{$^2$Department of Physics, University of Oxford, Keble Road, Oxford, OX1 3RH, U.K.}
\affil{$^3$Instituto de Astrof\'isica de Andalucia (CSIC), Granada, Spain}
\email{jorpega@mpia.de}

\begin{abstract}
We explore whether stellar tidal streams can provide information on the secular, cosmological evolution of the Milky Way's gravitational potential and on the presence of subhalos. We carry out long-term ($\Delta t\sim t_{\rm hubble}$) N-body simulations of disrupting satellite galaxies in a semi-analytic Galaxy potential where the dark matter halo and the subhalos evolve according to a $\Lambda$CDM cosmogony. All simulations are constrained to end up with the same position and velocity at present.
Our simulations account for: (i) the secular evolution of the host halo's mass, size and shape, (ii) the presence of subhalos and (iii) dynamical friction.\\
We find that tidal stream particles respond adiabatically to the Galaxy growth so that, at present, the energy and angular momentum distribution is exclusively determined by the present Galaxy potential. In other words, all present-day observables can only constrain the present mass distribution of the Galaxy independent of its past evolution.
We also show that, if the full phase-space distribution of a tidal stream is available,
we can accurately determine (i) the present Galaxy's shape and (ii) the amount of mass loss from the stream's progenitor, even if this evolution spanned a cosmologically significant epoch.
\end{abstract}


\section{Introduction}\label{sec:int}
In the last decade stellar streams in and around the Milky Way, which are possible debris from the disruption of satellite galaxies during the hierarchical assembly of our Galaxy, have become an active topic of investigation for several reasons. Firstly, large scale CCD surveys have provided unprecedented evidence of accretion and tidal disruption of dwarf galaxies around large
spirals in the Local Group (Milky Way: see Majewski 2004; M31: Ibata et al. 2002 and beyond: Pohlen et al. 2004). Secondly,  because tidal streams provide strong constraints on the potential of host galaxies, it is possible to estimate the shape of dark matter halos on large scales in contrast to traditional tracers, such as HI or stellar kinematics (see Sackett et al. 1999 for a review), which provide estimates on relatively small scales. The expected shape of dark matter halos depends on the nature of dark matter particles (see, for example, Dubinsky \& Calberg 1991, Yoshida et al. 2000 and Dav\'e et al. 2001 for shape estimates for cold, self-interacting and hot dark matter models, respectively), tidal streams represent a useful tool to discriminate between different paradigms. In addition, tidal stream properties also depend on the mass and internal structure of its progenitor (e.g Johnston, Sackett \& Bullock 2001 and Law, Johnston \& Majewski 2004, hereinafter LJM), which ultimately results in a complementary way of estimating the progenitor mass and therefore its mass-to-light ratio. Finally, tidal streams can be used to determine the position of the progenitor if this has not previously been detected (e.g. Font et al. 2004, Pe\~narrubia et al. 2005).

The formation of tidal streams is, conceptually, a simple process: along its orbit, a stellar system crosses regions where the tidal force of its host galaxy supplies kinetic energy to the initially bound particles. If the energy gain is large enough, particles can become unbound and escape from the host system, forming two sub-systems which are kinematically well differentiated: the leading and the trailing tails which, as their names indicate, precede and follow, respectively, the progenitor in its orbit. The orbital evolution of stripped particles is initially similar to that of the progenitor (e.g. Lynden-Bell \& Lynden-Bell 1995, Johnston 1998) although, as they evolve in the host galaxy potential, orbits diverge from each other monotonically with time. However, even after a large number of orbital periods, tidal stream particles and the progenitor system reside in well-defined regions of the constant of motion space (Helmi \& White 1999), therefore attesting a common origin.

The complexity of the stream formation and evolution forces the use of N-body calculations in most cases. The existing work can be divided into a) live N-body simulations, where the host galaxy is formed by a given number of particles initially in equilibrium and b) simulations where the host galaxy is represented by a non-responsive potential. 
Whereas the former takes into account the host galaxy's response to the satellite, the later neglects this in order to save computational resources for extensive orbit surveys. \\
However, none of the N-body simulations of tidal stream formation and evolution to date has accounted for the overall build-up of the host galaxy during the 1--10 Gyr that the stream formation may take. 
 Yet, 
in the commonly accepted hierarchical scenario, host galaxies experience large changes in mass, size and shape during their history and hence a tidal stream evolving in an unchanging host galaxy can only approximate recent epochs. 
The main goal of this contribution is to address the effect that the secular evolution of the host galaxy induces on the formation, evolution and interpretation of tidal streams. Specifically, we will explore whether the present-day structure of an extensive tidal stream can constrain the past history of the host's gravitational potential.

In addition to the secular overall mass and size growth of the host's halo, we will also examine the influence of dark matter substructures on tidal streams (Ibata et al. 2002 and Johnston, Spergel \& Haydn 2002). CDM cosmology predicts a large number of sub-structures in a galaxy-sized halo, many more than the number of observed dwarf galaxies (Klypin et al. 1999, Moore et al. 1999). Recently, it has been proposed that the process of re-ionization in the universe would lead to a significant decrease in the number of ``visible'' sub-structures (Bullock, Kravtsov \& Weinberg 2000, Benson et al. 2002, Somerville 2002, Tully et al. 2002), while keeping the total number of sub-structures unaltered. This process would establish a minimum mass (corresponding to a minimum circular velocity of $\sim 30$ km/s) above which gravitationally bound systems would be able to retain baryonic matter and, thus, to form stars. \\
The properties of kinematically cold tidal stream are strongly sensitive to the lumpiness of the galaxy potential (Ibata et al. 2002, Johnston, Spergel \& Haydn 2002) as repeated encounters with dark matter sub-structures alter the energy and angular momentum distribution of tidal stream particles, leading to hotter, broadly dispersed streams. Tidal streams appear to be a unique laboratory to determine the presence of dark matter clumps in galaxy halos. Firstly, because the effects of those clumps on tidal streams are purely gravitational (and so, independent of whether sub-structures enclose baryons or not) and, secondly, because tidal streams can be detected on large scales and can be as old as the host galaxy, and may therefore provide information on the number and spatial distribution of bound sub-structures at different epochs. Yet, the calculations mentioned above do not take into account the evolution of the spatial distribution nor the mass loss of substructures, which may weaken the influence of substructures on the tidal stream evolution.

In this paper we focus on tidal streams in the Milky Way simply because only for our Galaxy can streams be resolved into stars and accurate phase-space information gathered.

The remainder of this paper is arranged as follows. In \S\ref{sec:galmod} we describe our models for the Milky Way potential and for the satellite galaxies that are disrupted to form tidal streams; \S\ref{sec:code} details our N-body code used for evolving satellite orbits. \S\ref{sec:calc} describes the specific set of orbits we explored in this work. \S\ref{sec:mass} describes how we address the problem of the unknown mass loss history of the satellite. In \S\ref{sec:progorb} we explore the behavior of the satellite galaxy's orbit while in \S\ref{sec:tsprop} we explore the properties of the associated tidal streams. In \S\ref{sec:shape} we explore whether evolution of the halo shape can influence the properties of tidal streams. \S\ref{sec:clumps} examines the influence of dark matter substructures of these tidal streams.  Finally, in \S\ref{sec:disc} we present our conclusions.

\section{Models for the Milky Way and the Satellite Galaxies}\label{sec:galmod}

\subsection{Milky Way potential}
The host galaxy system is described by a time-dependent gravitational potential. Our Galaxy model consists of a Miyamoto-Nagai (1975) disc, a Hernquist (1990) bulge and a Navarro, Frenk \& White (1995, 1996, 1997) dark matter halo (hereafter, NFW halo). 

The gravitational potential of each of those components in cylindrical coordinates is:
\begin{eqnarray}
\Phi_d({\bf r})=-\frac{G M_d}{\sqrt{R^2+(a+\sqrt{z^2+b^2})^2}}, \label{eq:phid} \\
\Phi_b({\bf r})=-\frac{G M_b}{r+c},\label{eq:phib} \\
\label{eq:phih} \Phi_h({\bf r},t)=\frac{G M_h}{\ln(1+\frac{r_{\rm vir}}{r_s})-\frac{r_{\rm vir}}{r_s+r_{\rm vir}}}\frac{q}{2r_s}\times \\ \nonumber 
\bigg[\int_0^\infty\frac{m(u)}{1+m(u)}\frac{{\rm d}u}{(1+u)\sqrt{q^2+u}}-2\bigg],
\end{eqnarray}
where $\Phi_d({\bf r})$ and $\Phi_b({\bf r})$ are time-independent, but the halo potential $\Phi_h({\bf r},t)$ evolves as described below. There, $r_{\rm vir}, r_s$ are the virial and scale radii, respectively, $M_h=M_h(r_{\rm vir})$ and $r^2=R^2+z^2$. The potential of an axi-symmetric NFW halo was calculated from Chandrasekhar (1960) using elliptic coordinates
\begin{eqnarray}\label{eq:mcoord}
m^2(u)=\frac{R^2}{r_s^2(1+u)}+\frac{z^2}{r_s^2(q^2+u)},
\end{eqnarray}
where $q$ is the axis-ratio of iso-density surfaces. In this contribution, we shall denote {\it oblate} and {\it prolate} halos as those with $q<1$ and $q>1$, respectively. For spherical halos Eq.~(\ref{eq:phih}) reduces to
\begin{eqnarray}
\Phi_h=-\frac{G M_h}{\ln(1+\frac{r_{\rm vir}}{r_s})-\frac{r_{\rm vir}}{r_s+r_{\rm vir}}}\frac{\ln(1+r/r_s)}{r}\equiv \\ \nonumber
-V_c^2\frac{\ln(1+r/r_s)}{r/r_s}.
\label{eq:phih_sph}
\end{eqnarray}

Following Johnston et al. (1999) we fix the disc and bulge parameters as $M_d=1.0\times 10^{11}M_\odot$, $M_b=3.4 \times 10^{10} M_\odot$, $a=6.5 $kpc, $b=0.26$ kpc and $c=0.7 $kpc. The Milky Way halo parameters at $z=0$ were taken from Klypin, Zhao \& Somerville (2002) being $M_h=1.0\times 10^{12}M_\odot$, $r_{\rm vir}=258 $kpc, $r_s=21.5 $kpc, which leads to a concentration at the present epoch of $c=r_{\rm vir}/r_s=12$.

Additionally, we have defined a characteristic velocity, energy and angular momentum throughout this contribution, which have the values $v_{\rm ch}=262$ km/s, $E_{\rm ch}=v_{\rm ch}^2\simeq 68600$ (km/s)$^2$ and $L_{\rm ch}=4000$ kpc~km/s.

\begin{figure}
\plotone{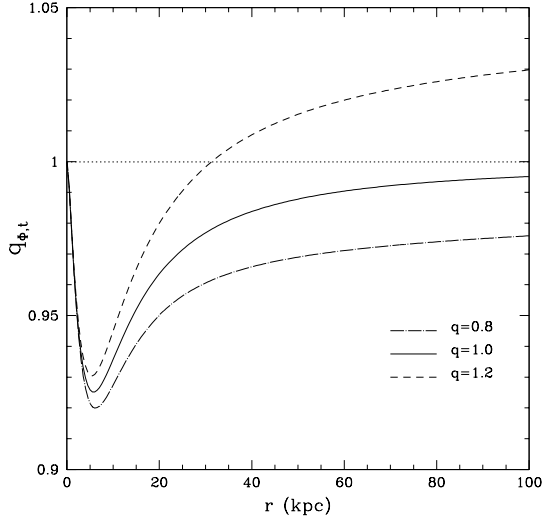}
\caption{Flattening of the total Galaxy potential as a function of Galactocentric distance for different halo axis-ratios. The horizontal dotted line indicates $q_{\Phi,t}=1$.}
\label{fig:qpot}
\end{figure}

Our selection of the different Milky Way components leads to a Galaxy potential that is aspherical everywhere. Note that for the gravitational potential at the present epoch, the only parameter that is varied in this work is the halo's density axis-ratio ($q$).
 In Fig.~\ref{fig:qpot} we plot the flattening of the {\it total} Galaxy potential as a function of distance and halo axis-ratio. One can clearly distinguish three regions: (i) the innermost region with $q_\Phi\simeq 1$ dominated by the spherical bulge component, (ii) an intermediate region, $3\leq r\leq 15$ kpc, where the disk potential dominates, so that the Galaxy potential has an oblate shape ($q_\Phi<1$) and (iii) the outer regions, $r>15$ kpc, where the halo potential dominates.

Some caveats must be kept in mind when using this simplistic Milky Way model:\\
Firstly, we have not included the evolution of the baryonic components. Observations at high redshift have shown that disc galaxies evolve in mass and size (e.g. Trujillo \& Pohlen 2005, Barden et al. 2005), which may induce effects similar to those of the varying halo potential on tidal streams. Moreover, the disc and bulge formation also changes the dark matter distribution, mostly in the inner halo region, as has been shown by several authors (e.g., Dubinsky 1994, Klypin, Zhao \& Somerville 2002). However, the self-consistent evolution of our Galaxy and its   effects on tidal streams would require the use of large N-body cosmological simulations, beyond the scope of this paper. Here, we shall merely examine tidal streams in a Milky Way-like galaxy, where only the dark matter component evolves.\\
Secondly, this evolution of our Milky Way model ignores any feed-back processes that satellite galaxies may induce on the host system.

\subsection{Average evolution of a Milky Way-sized halo}\label{sec:he}
In cold dark matter cosmogonies, halos grow hierarchically in mass and size owing to mergers of less massive systems. The process is stochastic in detail and is described by the merger tree of a halo. However, the average properties that determine the halo density profile at a given epoch have been studied exhaustively, both analytically and numerically, showing a smooth change with redshift or epoch. In this framework it is possible to trace back the evolution of a halo from its present density distribution.

 For the mean mass evolution Wechsler et al. (2002) find
\begin{equation}
M_h(z)=M_h(z=0) \exp\bigg[-a_c S\big (\frac{1}{a}-1 \big)\bigg],
\label{eq:mhevol}
\end{equation}
where $a=(1+z)^{-1}$, $a_c$ is the 'formation epoch' of the halo and $S=d\log M_h/d\log a$ is the log of the mass accretion rate at $a=a_c$. $S$ acts simply as a normalization constant in the fit of $M_h$; since its value can be arbitrarily chosen, we therefore fix $S=2$ as in Wechsler et al. (2002).\\
These authors also show that the average concentration of halos can be fit at any given time by the expression
\begin{equation}
c(z)=\frac{8.2}{S}\frac{a}{a_c}.
\label{eq:c}
\end{equation}
For a halo concentration of $c=12$ at $z=0$, for the Milky Way this fixes the formation epoch to $a_c\simeq 0.34$ (using S=2).

Bullock et al. (2001) have shown that the mean evolution of the virial radius can be expressed as
\begin{equation}
r_{\rm vir}(z)=\frac{75 {\rm kpc}h^{-1}}{1+z}\frac{M_h(z)}{1.0\times 10^{11} M_\odot h^{-1}} \bigg(\frac{200}{\Omega_0 \Delta_{\rm vir}(z)}\bigg)^{1/3},
\label{eq:rv}
\end{equation}
where $\Delta_{\rm vir}$ is the virial over-density, which can be written as (Bryan \& Norman 1998)
\begin{equation}
\Delta_{\rm vir}(z)=\frac{18\pi^2 + 82 [\Omega(z)-1] - 39 [\Omega(z)-1]^2}{\Omega(z)},
\label{eq:delv}
\end{equation}
 and $\Omega$ is the mass density of the Universe
\begin{equation}
\Omega(z)=\frac{\Omega_0 (1+z)^3}{\Omega_0 (1+z)^3+\Omega_\Lambda}.
\label{eq:omega}
\end{equation}
We have assumed for this study a $\Lambda$CDM Universe by  fixing $\Omega_0=0.3$, $\Omega_\Lambda=0.7$ and $h=0.7$.

Eq.~(\ref{eq:mhevol}),~(\ref{eq:c}) and ~(\ref{eq:rv}) determine the evolution of the halo potential $\Phi_h({\bf r},t)$ shown in Eq.~(\ref{eq:phih}).

\subsection{Statistical realizations of the Milky Way halo and the merger tree}\label{sec:mtree}
The stochastic growth of dark matter halos through mergers with smaller halos implies a tight correlation between the properties of the host halos and their merger trees.

In addition to the overall (smooth) mass growth outlined in Section~\ref{sec:he}, we also analyze the effects of dark matter sub-structures on tidal stream evolution by including an additional force component induced by these systems. 
We generate initial conditions for our calculations by constructing merger trees for an ensemble of overall identical dark matter halos
identified at $z=0$. 
Specifically, we take at $z=0$ dark matter halos of mass $1.0 \times 10^{12}M_\odot$, $r_{\rm vir}=258$ kpc and $c=12$. 
Following the method of Cole et al. (2000), we construct a statistical realization of the merging history of
these halos back to high redshift, assuming a $\Lambda$CDM cosmology with $\Omega_0=0.3$,
$\Omega_\Lambda=0.7$, $\sigma_8=0.9$, $h(=H_0/100\hbox{km s}^{-1}\hbox{ Mpc}^{-1})=0.7$ and $\Gamma=0.21$, and an
inflationary initial power spectrum (i.e. $P(k)\propto k$). At each point in time, we identify the most massive
progenitor of the $z=0$ halo and classify this as the ``host'' halo at that redshift. We then identify halos which are about to
merge with the host at each time and label these as new satellites.

Based on such merger trees, we record the virial mass, virial velocity and concentration of the host halo (assumed to have an
NFW density profile) at each epoch.
 We also record the same information for each satellite along with the epoch of each merging event. 
The initial orbit for each satellite is set by choosing an initial position at
random on the surface of a sphere with radius equal to $r_{\rm vir}(z)$ of the host halo, while the initial velocity is
chosen from the distribution found by Benson (2004) for $z=0$ in $\Lambda$CDM cosmologies\footnote{Benson (2004) finds some
evidence for evolution of the velocity distribution to higher redshifts, and for dependence on the masses of the merging
halos. These dependencies are poorly quantified, and so we ignore them here.}.

In this way we generate both the formation history of the host halo
and the properties of the infalling population of satellites. These
are used as inputs to our model for the evolution of substructures (Pe\~narrubia \& Benson 2005). The outcome is nine different merger trees of a Milky Way-like galaxy for which, at each time-step, we have the position, velocity, mass, virial and scale radii of all subhalos with masses above $10^7 M_\odot$.

\begin{figure}
\plotone{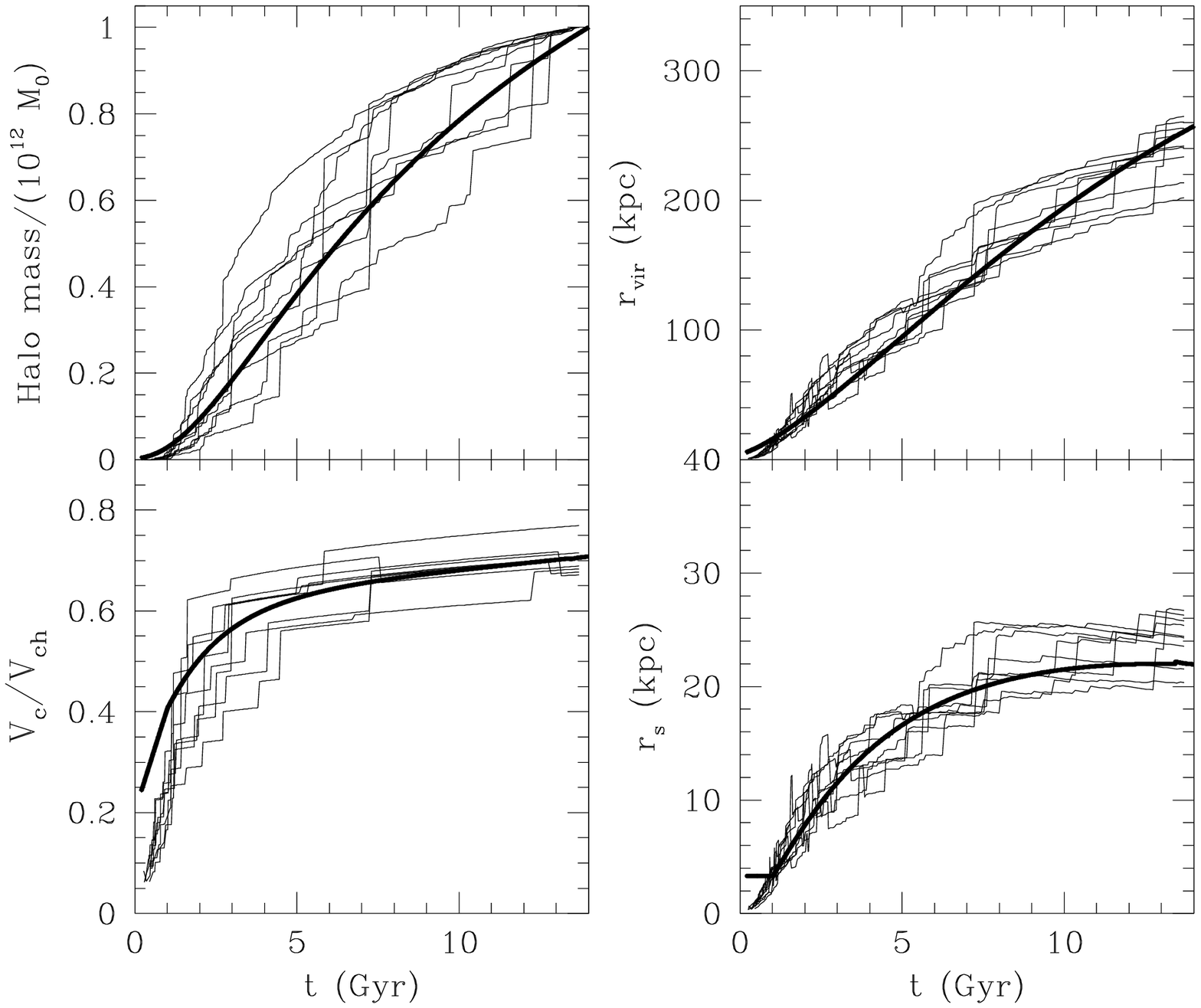}
\caption{Evolution of the dark matter properties for the host halo in our simulations. {\it Upper-left panel:} Evolution of the halo mass enclosed within the virial radius. {\it Upper-right panel:} Evolution of the halo virial radius. {\it Bottom-right panel:} Scale-radius evolution. {\it Bottom-left panel:}Halo circular velocity evolution (see text for the definition of $V_c$ as a function of $M_h, r_{\rm vir}, r_s$).   Thin lines show nine statistical realizations of the merger tree of a Milky Way-like galaxy. Thick lines show the mean evolution of the Galaxy halo as outlined in Section~\ref{sec:he}.}
\label{fig:r}
\end{figure}

\subsection{Satellite models}\label{sec:satmod}
The actual disruption of satellites is modeled by `live' (i.e. self-consistent, self-gravitating) N-body realizations of satellites in an evolving, but non-responsive, halo potential. We also include perturbations from the ensemble of dark matter subhalos.

We have realized our numerical models of satellite galaxies with King profiles (King 1966), which approximate the light profile of the Galaxy's low-mass satellites (e.g. Irwin \& Hatzidimitriou 1995). The high mass-to-light ratios measured in dwarf galaxies (such as Ursa Minor and Draco, see Mateo 1998, Kleyna et al. 2004, Locas, Mamon \& Prada 2005, Mu\~noz et al. 2005) indicate that these systems are dark matter dominated , but with a dark matter profile that is not well determined.\\
As Zhao (2004) showed, the mass and orbital evolution are fairly sensitive to the specific mass profile we assume for our satellites, an aspect that must be kept in mind, but it is not pursued further here.

For the initial satellite structure we assume a King profile with a dimensionless central potential $W_0=4$, or concentration parameter $c\equiv\log_{10}(r_t/r_k)\simeq 0.84$, where $r_k$ and $r_t$ are the King and tidal radii, respectively. Our satellite models have $10^5$ particles. 
The initial mass and tidal radius are chosen such that the mass at the present epoch is $M_s(t=t_f)=5\times 10^8 M_\odot$ in all of our calculations. Details of how to make this choice are given in Section~\ref{sec:calc}.

\section{Modeling the disruption of N-body satellite galaxies}\label{sec:code}
The force acting on each satellite particle in our simulations has four components: 1) the gravity of all other satellite particles; 2) the force from the smooth, non-responsive Milky Way potential; 3) dynamical friction on the still bound satellite portion and 4) the time-dependent forces from subhalos.

 To calculate the force from self-gravitating satellite particles we use {\sc
Superbox}, which is a highly efficient particle mesh-algorithm based on
a leap-frog scheme (for details see Fellhauer et al. 2000).

The acceleration from the Milky Way gravitational potential is calculated from Eq.~(\ref{eq:phid}), (\ref{eq:phib}) and (\ref{eq:phih}) and included in the N-body code as an external force.

The code explicitly includes the dynamical friction force term resulting from the interaction of the satellite galaxy with the smooth dark matter component. In axi-symmetric self-gravitating halos, the presence of a satellite galaxy traveling through the system induces a density wake behind the satellite, which exerts a drag force on the satellite particles. Following Binney (1977) this can be approximated by
\begin{eqnarray}\label{eqn:ffbinn}
f_{ i,df}=-\frac{\sqrt{2\pi}\rho_h[m^2(0)] G^2M_s \sqrt{1-e_v^2} {\rm
ln}\Lambda}{\sigma^2_R \sigma_z}B_R v_{i}, \\ \nonumber
f_{ z,df}=-\frac{\sqrt{2\pi}\rho_h[m^2(0)] G^2M_s \sqrt{1-e_v^2}{\rm
ln}\Lambda}{\sigma^2_R\sigma_z}B_z v_{z},
\end{eqnarray}
\noindent where $i=x,y$ and $(\sigma_R,\sigma_z)$ is the velocity dispersion
ellipsoid in cylindrical coordinates with constant ellipticity
$e_v^2=1-(\sigma_z/\sigma_R)^2$. Here $B_R$ and $B_z$ are given by
$$ B_R=\int_0^\infty dq \frac{\rm{exp}
(-\frac{v_{R}^2/2\sigma^2_R}{1+q}-\frac{v_{z}^2/2\sigma^2_{R}}{1-e_v^2+q})}
{(1+q)^2(1-e_v^2+q)^{1/2}}, $$
$$ B_z=\int_0^\infty dq \frac{\rm{exp}
(-\frac{v_{R}^2/2\sigma^2_R}{1+q}-\frac{v_{z}^2/2\sigma^2_{R}}{1-e_v^2+q})}
{(1+q)(1-e_v^2+q)^{3/2}}, $$ 
with $(v_R,v_z)$ is the satellite velocity in this frame and $\ln\Lambda$ the Coulomb logarithm. 

The NFW density profile in elliptical coordinates can be written as
\begin{equation}\label{eq:rhoh}
\rho_{h}=\frac{M_h}{4\pi r_{\rm s}^3\big[\ln(1+r_{\rm vir}/r_{\rm s})-r_{\rm s}/(r_{\rm s}+r_{\rm vir})\big]}\frac{1}{m(0)[1+m(0)]^2}
\end{equation}
where we have that $m^2(0)=(R^2+z^2/q^2)/r_s^2$ from Eq.~(\ref{eq:mcoord}).

At each point in time, the satellite mass $M_s$ is defined as the sum of still-bound particles (see Section~\ref{sec:mass}).

 Pe\~narrubia, Just \& Kroupa (2004) have checked these equations against several satellite orbits in self-consistent axi-symmetric systems by keeping the Coulomb logarithm as a free parameter. Best fits were found for $\ln\Lambda=2.1$, almost independent on the satellite's mass, orbital eccentricity and halo axis-ratio.

In order to implement Eq.~(\ref{eqn:ffbinn}), we have assumed that only bound particles feel dynamical friction, whereas unbound (stripped) particles do not. In this way we are neglecting the effects that the self-gravity of the density wake may induce on the orbit of the escaping particles. This appears a sensible approximation as most of the unbound particles remain close to the main system (and hence the density wake) for only a short time and, furthermore, that $f_{\rm df}\sim [10^{-3},10^{-2}] d\Phi_r/dr$ (Just \& Pe\~narrubia 2005).

The last force component results from the presence of dark matter clumps. Using the semi-analytic algorithm presented in Pe\~narrubia \& Benson (2005), we calculate and store the position, velocity and internal properties of dark matter sub-structures at each time-step. This semi-analytic code has been proved to reproduce fairly well the mass and spatial distributions of dark matter sub-structures obtained from large cosmological N-body simulations. In each step of the satellite simulation, the force on the $i^{\rm th}$ from all subhalos is
 \begin{equation}\label{eq:f_sub}
{\bf f}_{i,\rm sub}=\Sigma^{N_{\rm sub}}_{j=1} G M_j\frac{{\bf r}_j-{\bf r}_i}{(\epsilon_j^2+ | {\bf r}_j-{\bf r}_i  |^2)^{3/2}}
\end{equation}
where $\epsilon_j$ is the {\em softening parameter} of the $j^{\rm th}$ substructure which accounts for its finite extent. We assume that $\epsilon_j=r_{h,j}$, where $r_{h,j}$ is the half-mass radius.

We must remark that, in contrast to previous studies of the effects of halo lumpiness on tidal streams (Ibata et al. 2002, Johnston, Spergel \& Haydn 2002), our approach accounts for the evolution of mass, spatial distribution and size of dark matter sub-halos.

After calculating all the force terms, {\sc Superbox} solves the equation of motion of each satellite particle
 \begin{equation}\label{eq:eqmot}
\frac{d^2 {\bf r}_i}{d t^2}=-{\bf \nabla}(\Phi_s+\Phi_d+\Phi_b+\Phi_h)_i + \xi {\bf f}_{i,\rm {df}}  + {\bf f}_{i,\rm{sub}}
\end{equation}
where  $\Phi_s$ is the self-gravitational potential of the satellite galaxy calculated by {\sc superbox}, $\xi=0$ for unbound particles and $\xi=1$ for bound ones.
{\sc Superbox} uses a leap-frog scheme with a constant time-step of $\Delta t=0.65$ Myr, which is about $1/100$th the
dynamical time of our most massive satellite model. We have three resolution zones, each
with $64^3$ grid-cells: (i) The inner grid covers out to 10 King radii, providing a resolution of $r_k/6$. (ii) The
middle grid covers 20 King radii $r_k/3$. (iii) The outermost grid extends to 348~kpc
and contains the local universe, at a resolution of 11.6 kpc. At any time-step, all grids are centered at the density maximum of the satellite model. 

This choice of grids provides maximal resolution within the satellite itself, where the self-gravity of the satellite dominates. Most unbound particles eventually reach $r>20 r_k$ from the satellite center, i.e,  within the lowest-resolution zone. This, however, induces negligible effects, for the orbit of stripped particles is mainly determined by the semi-analytical Galaxy potential (and so is insensitive to spatial-resolution problems).

\begin{table*}
\begin{center}
 \caption{Models}\label{tab:calc}
\begin{tabular}{||l |l |l  |l |l |l  |l |l |l   |l |l |l  |l ||}\tableline \tableline
{\bf Name} & M1H1Q1 & M1H1Q2 & M1H1Q3 &  M1H2Q1 & M1H2Q2 & M1H2Q3 &  M2H1Q1 & M2H1Q2 & M2H1Q3 & M2H2Q1 & M2H2Q2 & M2H2Q3 \\\tableline
$M_s(t_0)/M_s(t_f)$ (1)& 2 &  2  & 2 & 2 & 2 & 2 & 10 & 10 & 10 & 10 & 10& 10\\
$r_k$ (kpc) (2) & 0.55 & 0.55 & 0.55 & 0.58 & 0.58 & 0.58 & 1.20 & 1.20 & 1.20 & 1.30 & 1.30 & 1.30 \\
$q$ & 0.8 & 1.0 & 1.2 & 0.8 & 1.0 & 1.2 & 0.8 & 1.0 & 1.2 & 0.8 & 1.0 & 1.2 \\
Halo (3) & static & static & static & evolving & evolving &evolving & static & static & static &evolving & evolving &evolving \\ \tableline
\end{tabular}
\end{center}
\tablenotetext*{}{(1) We fix $t_0=4$ Gyr and $t_f=t_{\rm Hubble}=14$ Gyr. All satellite models have a final time mass of $M(t_f)=5\times 10^8 M_\odot$.\\ 
(2) King radius at $t=t_0$. For $W_0/\sigma^2=4$ the tidal radius is $r_t=6.9 r_k$.\\
(3) Static halos have $M_h(t)=M_h(t_f)$, $r_s(t)=r_s(t_f)$ and $r_{\rm vir}(t)=r_{\rm vir}(t_f)$. }
\end{table*}

\section{Calculations}\label{sec:calc}
With the modeling tools at hand we now explore what can be learnt about the dynamical history of a disrupting satellite and its tidal tail on the basis of present-day observables. One additional open point addressed in this Section is how to reconstruct the mass-loss history and the orbit of a satellite galaxy using an iterative method.

\subsection{Present-day parameters}
We have focused our study on a Sagittarius-like galaxy because, at the present day, it is the system in the Milky Way for which the most comprehensive observational constraints exist. Following LJM (and references therein) we adopt the following remnant properties at $t=t_f=14 $Gyr: 

\begin{itemize}
\item $M_s(t_f)=5\times10^8 M_\odot$;
\item $r(t_f)=r_{\rm peri}=15 $kpc;
\item $v_{\rm tan}(t_f)=320$ km/s;
\item Orbital inclination $i=45^\circ$.
\end{itemize}

With the \emph{present} mass, position and velocity of a satellite galaxy fixed, we explore the physical processes that may have influenced the orbit of the satellite galaxy \emph{in the past}:\\
1) Axis-ratio of the host halo.\\
2) Dynamical friction.\\
3) Smooth evolution of the Galaxy's potential.\\
4) The presence of a large number of dark matter sub-structures.

To analyze the effects of various factors on the satellite orbit and stream evolution we have conducted several N-body simulations. Table~\ref{tab:calc} lists the properties of the satellite and host halo used in each simulation.
\vskip0.3cm

\subsection{Iterative reconstruction of the satellite orbit}
Each model has been evolved in the following way:\\
(i) The present time is denoted as $t_f=14$ Gyr. Starting from the present, we integrate models {\it backwards} in time to $t_0=4$~Gyr~
\footnote{In order to fix $t_0$ we have applied the results of Zentner \& Bullock (2003) and Pe\~narrubia \& Benson (2005). These authors show that, as a result of tidal disruption, the maximum accretion time of sub-structures with $M(t_f)>10^8 M_\odot$  is $t_f-t_0\simeq 10$ Gyr. } 
by solving Eq.~\ref{eq:eqmot} with $\xi=-1$ and ${\bf f}_{\rm sub}=0$ for a single particle. The goal is to determine ${\bf r},{\bf v}$ at $t=t_0$ for a given mass loss history. \\

(ii) Using the results of (i), we place each N-body satellite realization at ${\bf r}(t_0)$ with velocity $-{\bf v}(t_0)$. Subsequently, we evolve the system {\it forwards} in time using the N-body algorithm (\S\ref{sec:code}), selecting those realizations which produce N-body satellites that match our constraints at $t=t_f$ .

This method assures that all models match the present constraints independently of their past evolution. However, point (i) cannot be trivially solved since the solution of the equation of motion (Eq.~\ref{eq:eqmot}) is coupled to that of the mass evolution through the dynamical friction term (which scales as $\rho_g({\bf r}) M_s(t)$, where $\rho_g$ is the Galaxy density profile). In practice, that means that  the satellite's orbit and mass must be simultaneously calculated {\it backwards} in time. 

\subsection{Modeling satellite mass loss}\label{sec:mass}
In order to disentangle the past mass and orbit evolution we have applied the following scheme:\\
The fist step is to fix the total amount of mass loss. Observational data indicate that satellite galaxies appear dark matter dominated (e.g. Mateo 1998), which precludes a reconstruction of the total mass loss history from the present-day distribution of stars within a satellite galaxy. Therefore, we treat the total amount of mass loss as a free parameter. According to Pe\~narrubia \& Benson (2005): (i) The present-day population of subhalos have lost on average 50\% of their mass since they were accreted and (ii) mass loss events where subhalos lose more than 90\% of their accretion mass are extremely rare. To analyze typical and extreme cases we carry out simulations where $M_s(t_0)=2M_s(t_f)=10^9 M_\odot$ (models M1) and $M_s(t_0)=10M_s(t_f)=5\times 10^9 M_\odot$ (models M2, see Table~\ref{tab:calc}).

 The second step is to determine the mass loss history, i.e, $M_s(t)$. Here we use the results of Zhao (2004), who showed that, for a given satellite density profile, the mass evolution of satellite galaxies can be described by a set of empirical functions. In particular we use
 \begin{eqnarray}\label{eq:m}
\frac{M_s(t)}{M_s(t_f)} &= & \frac{M_s(t_0)}{M_s(t_f)} \bigg[1- \left\{\exp\bigg(-\frac{t_f-t}{t_f}\bigg)\right\}^p \nonumber \\
&& +\left\{\exp\bigg(-\frac{t_f-t_0}{t_f}\bigg)\right\}^p\bigg]
\end{eqnarray}
where the power $p$ depends on the mass loss fraction $M_s(t_f)/M_s(t_0)$. The time interval is $t\in[t_0,t_f]$ with $t_f\leq t_0$.

 Lastly, Eq.~(\ref{eq:m}) still requires to fix the value $p$ for each simulation. Since orbit and mass are uniquely related once $M_s(t_0)/M_s(t_f)$ has been chosen, we use the following iterative scheme to determine $p$ and solve Eq.~(\ref{eq:eqmot}) for each satellite model: 

0) {\it Selecting the starting $p$-value:} Assuming $M_s(t)={\rm const}.$, we integrate the orbit backwards in time. Subsequently, we run the N-body orbit forwards in time in order to calculate the number of bound particles as a function of time and to fit an initial, tentative value of $p$.\\
1){\it Orbit calculation}: With the tentative $p$-value we integrate Eq.~(\ref{eq:eqmot}) backwards in time with $M_s(t)$ given by Eq.~(\ref{eq:m}).\\
2){\it Fixing the initial satellite size and the total mass loss}: We have used the fact that the initial satellite's tidal radius and its mass loss rate are correlated. 
Using the mass evolution and the orbit obtained in (i), we run several N-body simulations varying the initial $r_{\rm t}$ (see \S\ref{sec:satmod}) with a fixed $M_{\rm s}(t_0)$ and select those models that end up with $M_s(t_f)\simeq 5\times 10^8 M_\odot$. \\
3) {\it Fine-tuning}: Finally, we fine-tune the value of $p$ to obtain a better match the the mass-loss histories derived from the N-body simulations. 

This process repeats until the N-body and the empirical $M_{\rm s}(t)$ match\footnote{Note that $M_s(t_f)$ does not enter into Eq.~(\ref{eq:m}), and so the present mass depends only on the value of $p$. Since $p$ is found from the best fit to the mass curve within $[t_0,t_f]$, Eq.~(\ref{eq:m}) may, for large values of $M_s(t_0)$, give a slightly different value for $M_s(t_f)$ than desired. We have checked that this small mismatch induces negligible effects on the orbit integration.}, typically 2--3 iterations.

 \begin{figure}
\plotone{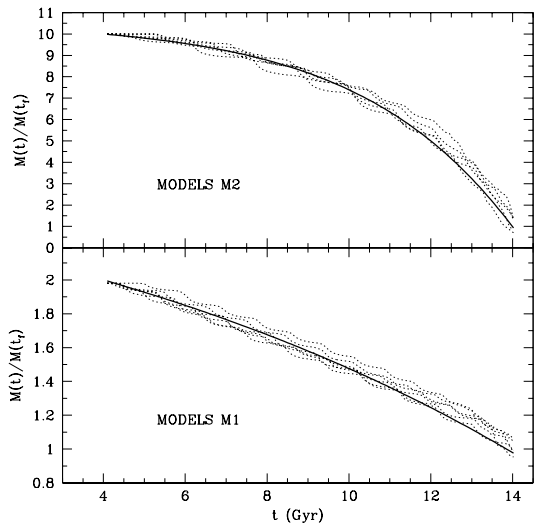}
\caption{Mass evolution of our satellite models. Upper and lower panels show the mass loss of satellites with initial masses $M_s(t_0)=5\times 10^9 M_\odot$ (family model M2) and $M_s(t_0)=10^9 M_\odot$ (family model M1), respectively. All models have the same mass at present ($M_s[t_f]=5\times 10^8 M_\odot$). Thick lines show the curves obtained from Eq.~(\ref{eq:m}). Dotted lines show the fraction of bound particles obtained from the self-consistent satellite potential calculated by {\sc superbox} for each satellite of the same family model. Note that the mass loss curves only depend on $M_s(t_0)/M_s(t_f)$ and not on the exact Galaxy potential.} 
\label{fig:mass}
\end{figure}

In Table~\ref{tab:calc} we have listed the properties of our satellite models and in Fig.~\ref{fig:mass} we show the mass evolution of satellites with $M_s(t_0)/M_s(t_f)=10$ (upper panel, family models M2) and $M_s(t_0)/M_s(t_f)=2$ (lower panel, family models M1). Thick lines show the empirical evolution obtained from Eq.~(\ref{eq:m}) with $p=1$ (lower panel) and $p=4$ (upper panel). Dotted lines show the amount of bound mass calculated from the N-body models as a function of time: The number of bound particles at any time-step is determined from the self-consistent potential calculated by {\sc superbox}. We label a particle as unbound whenever $E=\frac{1}{2}({\bf v}-{\bf v_{cm}})^2+\Phi_s>0$, where ${\bf v_{cm}}$ is the center-of-mass velocity of the bound particles, ${\bf v}$ is the particle velocity in the Galaxy frame and $\Phi_s$ is the potential induced by the satellite's self-gravity.

Interestingly, we find that the shape of the mass loss function {\it only} depends on the total amount of mass loss and {\it not} on the dark matter halo's parameters, such as mass, size and density axis-ratio, so that once $M_s(t_0),M_s(t_f),({\bf r_s,v_s})(t_f)$ are fixed, there exists a unique value $r_t(t_0)$ that reproduces the function $M_s(t)$ shown in Eq.~(\ref{eq:m}).

As commented above, the exact value of $r_t(t_0)$ and $M_s(t_0)/M_s(t_f)$ are directly correlated. As we show in Table~\ref{tab:calc}, satellite models in evolving halos (models H2) have initial tidal radii a factor 1.06 larger than those models in static halos (models H1), which indicates that mass loss is slightly reduced if halo evolution is implemented.
The relation between the initial tidal radius and the mass loss fraction is more evident if comparing models M1 against models M2. There we see that strong mass loss events imply large initial tidal radii. In particular we find that $r_{t,M2}/r_{t,M1}\simeq 2.2$ for $M_{s,M2}/M_{s,M1}=5$ at $t=t_0$.

Although this technique is computationally expensive (for every simulation one must run several N-body simulations in order to fix $r_t(t_0)$ and $p$), it avoids the use of complex semi-analytic algorithms to estimate  mass evolution (e.g. Taylor \& Babul 2001, Zentner \& Bullock 2003, Pe\~narrubia \& Benson 2005) which are considerably less accurate.

\section{Satellite orbit }\label{sec:progorb}
Here we describe the main factors that determine the long-term orbit evolution of the satellite up to the present.

\subsection{Dependence on the Halo Axis-ratio} 
As already mentioned in the Introduction, the Galaxy's halo axis-ratio has not yet been fully constrained. In practice that means that, whereas the present angular momentum can be directly measured from observations, the orbital energy has a range of possible values depending on the halo's axis-ratio ($q$).

\begin{figure}
\vspace{5.5cm}
\includegraphics{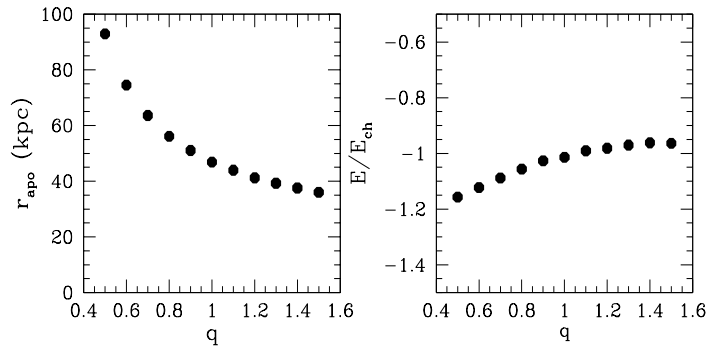}
\caption{Left panel: First apo-center distance as a function of the halo axis-ratio ($q$). Right panel: Initial energy as a function of $q$. In all calculations we fix the present satellite position at $r=15$ kpc (peri-center) with a tangential velocity of $v_{\rm tan}=320$ km/s. }
\label{fig:rapoe}
\end{figure}

In Fig.~\ref{fig:rapoe} we show the value of the energy at $t=t_f$ as a function of $q$ (right panel) and the resulting last apo-center distance ($t\simeq t_f$, left panel). As we can see, the apo-center is fairly sensitive to the halo axis-ratio, specially for $q<1$, which allows this last quantity to be constrained from the spatial distribution of the ``young'' tidal stream\footnote{As an example, LJM have used the radial extension of the Sgr stream to perform a similar study. Unfortunately, no accurate proper motions of the Sgr main system are available, so that $v_{\rm tan}$ was a free parameter in their models.}. In Section~\ref{sec:tsprop} we shall discuss other constraints on $q$ from long-term orbital properties (i.e, from older stream pieces).

\subsection{Satellite orbit and the Mass-Loss History}
The orbit and the mass of satellite galaxies are tightly correlated through dynamical friction. 

In Fig.~\ref{fig:rq10} we plot the galactocentric distance as a function of time in the absence of dynamical friction (upper panel) and for two different mass loss histories: Models M1 (middle panel) and M2 (bottom panel). For clarity, we only show orbits in a spherical halo. Solid and dotted lines denote, respectively, orbits in static and evolving halos. \\
Looking at the orbits in static halos, we find that dynamical friction decreases both the peri- and apo-center distances, which for the model M1 is approximately $r_{\rm apo}(t_0)/r_{\rm apo}(t_f)\simeq 1.33$ and for the model M2 $r_{\rm apo}(t_0)/r_{\rm apo}(t_f)\simeq 2$ (similar values are found for $r_{\rm peri}$). Note that the orbit of the model M2 decays only 1.5 times more than M1, even though its mass loss is fivefold.
The fact that the orbital decay is not simply proportional to the satellite mass can be readily shown by a dimensional calculation: taking into account that $f_{df}\propto M_s\rho\sim M_s/r^3$, we have that $f_{df,M2}/f_{df,M1}\sim M_{s,M2}/M_{s,M1} r_{\rm apo,M1}^3/r_{\rm apo,M2}^3\simeq 1.48$ 
at $t=t_0$ (note that this value is $f_{df,M2}/f_{df,M1}=1$ at $t=t_f$) which indicates that dynamical friction rapidly decreases as the averaged galactocentric distance increases.

\subsection{Satellite orbit and the Evolution of the Milky Way Halo}
\begin{figure}
\plotone{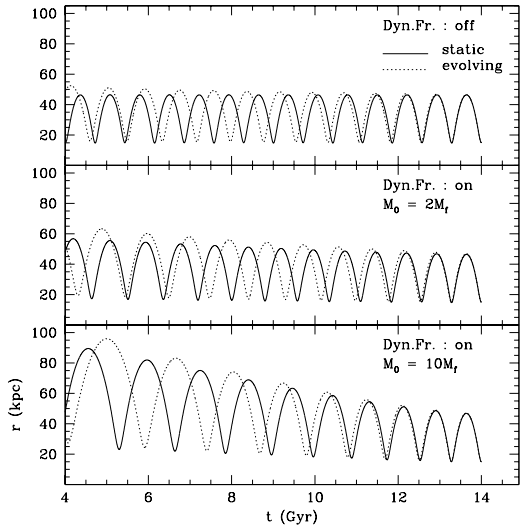}
\caption{Galactocentric distance as a function of time. In the upper panel we show the satellite's orbit in absence of dynamical friction. In the middle and bottom panels we show, respectively, the orbits of the models M1 and M2. Solid and dotted lines denote orbits in static and evolving halos, respectively. The axis-ratio is fixed to $q=1$. Note that all models end up at present ($t=14$ Gyr) at the same position and with the same velocity.}
\label{fig:rq10}
\end{figure}

In Fig.~\ref{fig:rq10} we also plot the orbit evolution of the models M1 and M2 in an evolving halo (dotted lines). Because the halo was smaller and less massive in the past, the apo and peri-center distances were larger at earlier times than at $t\simeq t_f$, even in the absence of dynamical friction (upper panel). Comparing orbits in static and evolving halos, we find that the maximum increase in the apocentric distance (i.e. at $t=t_0$) is approximately 10\%, independent of the satellite mass.

\begin{figure}
\plotone{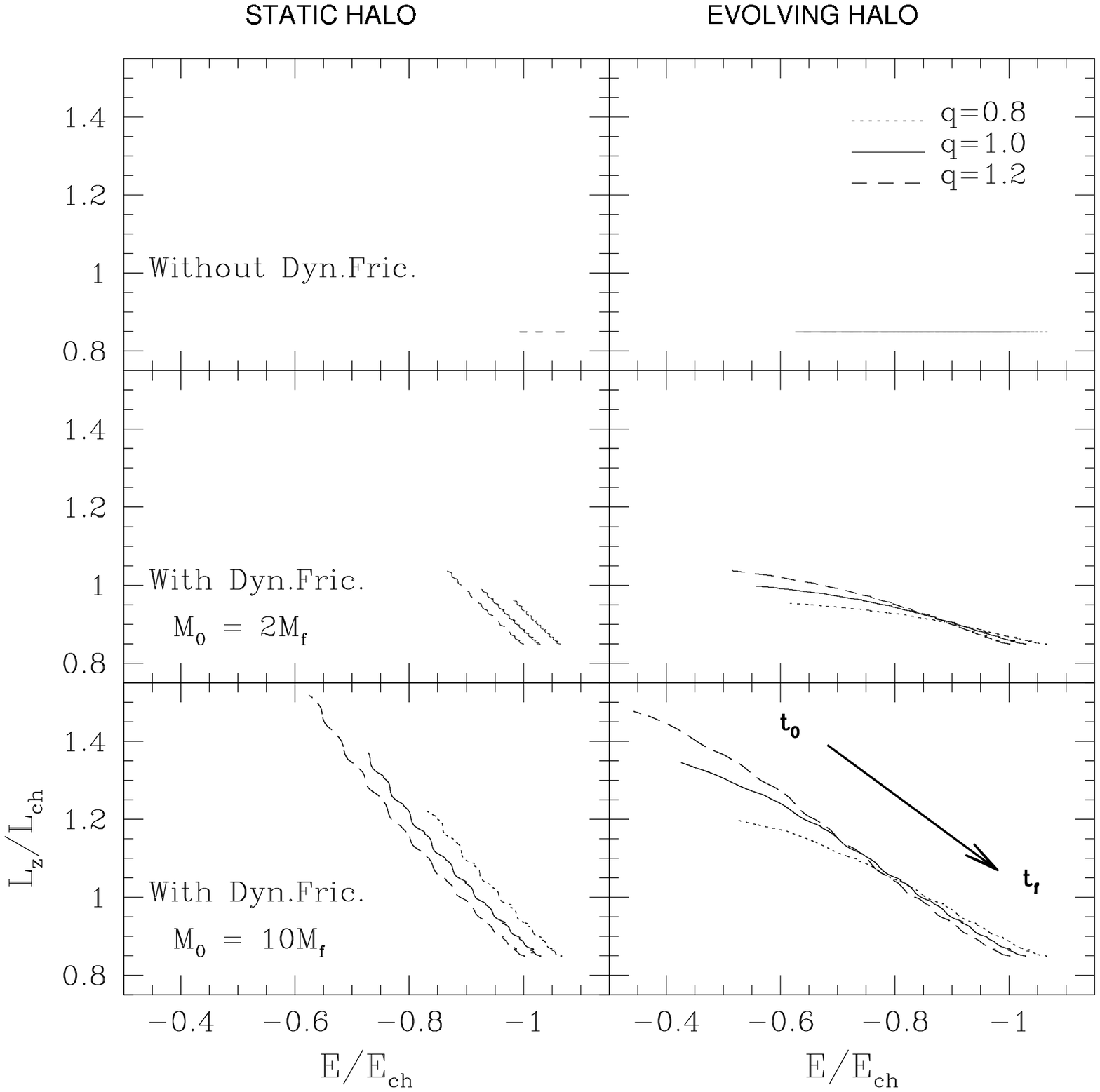}
\caption{Evolution of the satellite's orbit in the constant-of-motion plane. Upper, middle and bottom raws show, respectively, orbits in the absence of dynamical friction and the orbits of models M1 and M2. Dotted, full and dashed lines show orbits in halos with axis-ratios of $q=0.8,1.0, 1.2$, respectively. The arrow in the bottom-right panel indicates the direction of the time evolution. By construction, all models have the same mass, position and velocity at $t=t_f$.}
\label{fig:le}
\end{figure}

Although, at first glance, halo evolution and dynamical friction appear to induce similar effects, the orbital properties evolve in quite different ways.
In Fig.~\ref{fig:le} we plot the evolution of the energy and angular momentum in a static and an evolving Galaxy potential for the models shown in Table~\ref{tab:calc}. \\
The upper row shows orbits in the absence of dynamical friction. Without dynamical friction in a static halo, the energy and the perpendicular component of the angular momentum ($L_z$) are constants of motion, so that a given orbit is represented by a point in the $E-L_z$ plane\footnote{In practice, we see a short, horizontal line in the figure as energy is not perfectly conserved in our N-body scheme.}. The different energy values for $q=0.8, 1.0, 1.2$ correspond to those plotted in Fig.~\ref{fig:rapoe}.\\
 In contrast, if the halo potential evolves, $L_z$ is an adiabatic invariant (see \S3.6 of Binney \& Tremaine 1987) while the orbital energy increases (note that $E<0$) as we integrate the orbit backwards in time, resulting in an increase of the peri and apo-centric distances shown in Fig.~\ref{fig:rq10} (upper panel).\\
If dynamical friction is switched on (middle and bottom rows) both, $E$ and $L_z$, had higher values in the past. Comparing curves with different halo axis-ratios, we find that, independent of the satellite mass, the secular shift of energy and angular momentum is larger in prolate halos ($q>1$) than in oblate ($q<1$) ones. 

Therefore, we conclude that, even without dynamical friction, the halo growth does not alter the momentum evolution, as this last quantity is an adiabatic invariant, although the energy variation increases by a factor $\simeq 1.7$ for all satellite models, independently of satellite mass and halo axis-ratio.

\section{Tidal stream properties}\label{sec:tsprop}
In Section~\ref{sec:progorb} we have analyzed the orbits of satellite galaxies under different mass loss histories and halo potentials, with the only constraint that the present mass, velocity and position is the same in all models. In this Section we use N-body simulations to explore the properties of the tidal streams associated with these satellite galaxies. The goal is to determine how much information on the halo potential and on the satellite mass history can be obtained from tidal streams.

\subsection{The halo flattening and the orbital plane evolution}\label{sec:orbplane}
At any instant, the angular momentum ${\bf L}$ of a particle orbiting in a gravitational potential corresponds to the normal vector of the orbital plane.
Therefore, in spherical coordinates, the orbital plane can be described by two angles, the azimuthal angle $\phi\equiv \tan^{-1}(L_y/L_x)$ and the  orbital inclination $i\equiv \cos^{-1} (L_z/|L|)$. If the Galactic potential is not spherical and $i\neq0^\circ$ and $i\neq 90^\circ$, we expect the orbital plane to {\it nutate} ($di/dt\neq 0$) and to {\it precess} ($d\phi/dt\neq 0$).

As Johnston, Law \& Majewski (2005) have shown, determining the precession rate of tidal streams provides a powerful technique to measure the Galactic shape since both effects, precession and nutation, are induced by the quadrupole of the total Galaxy potential.  Moreover, this method does not require any kinematical information, since the normal vector can be obtained by fitting the 3D distribution of debris onto a plane.

\begin{figure}
\plotone{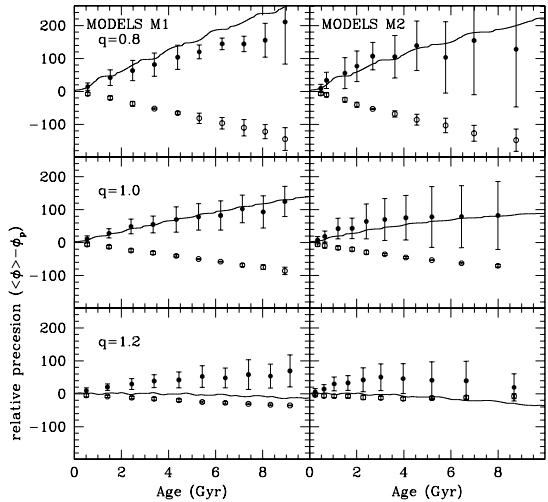}
\caption{Precession angle ($\phi$) between the tidal debris and the parent satellite at $t=t_f$ as a function of the stream age, for different halo axis-ratios and initial ($t=t_0$) satellite masses. Filled and and open symbols denote particles in the leading and trailing tails, respectively. $\phi_P$ is the present azimuthal angle of the stream's progenitor. Solid lines show the progenitor's precession. These models assume a static halo (models H1).}
\label{fig:phi}
\end{figure}

In Fig.~\ref{fig:phi} we show the evolution of the azimuthal angle as a function of the stream age\footnote{We define ``age'' of a stream particle as $t_f-t_u$, where $t_u$ is the time when the particle became unbound and $t_f=14$ Gyr (present).} for the models H1 (static halo, see Table~\ref{tab:calc}). We have divided the stream particles of the present epoch into leading and trailing tail particles (filled and open symbols, respectively) for different age sub-samples. Subsequently, we calculate $\langle \phi\rangle$ and the mean variance (this last represented by error bars). Additionally, we have plotted the evolution of the progenitor's azimuthal angle (full lines). This Figure shows that:\\
(i) Tidal streams do not present a single orbital plane but two, corresponding to the trailing and leading tails.\\
(ii) As Johnston et al. (2005) finds, the precession rate $d\phi/dt$ has a different sign for leading and trailing tail planes. In particular, leading tail planes precess as one would expect in an oblate galactic potential, whereas trailing tail planes precess in the opposite sense. \\
(iii) The precession rate is sensitive to the total amount of mass loss. Satellite models that have lost a large amount of the initial mass show a smaller precession rate in the oldest parts of the stream in comparison with satellite galaxies that were originally less massive. That reflects the fact that, as a result of a larger dynamical friction, models M2 moved at larger distances than models M1 (see Fig.~\ref{fig:rq10}), in regions where the Galaxy potential is more spherical (see Fig.~\ref{fig:qpot}).\\
(iv) Leading tails show a larger dispersion around the mean precession angle than trailing tails. We can also see that the variance (shown with error bars and defined as $\langle (\phi-\langle\phi\rangle)^2\rangle^{1/2}$) of the leading tails increases with the initial satellite mass and with the stream age. \\
(v) The evolution of the satellite's azimuthal angle is fairly similar to that of the leading tail if the Galaxy halo is oblate. If the halo is prolate, the precession rate has opposite sign, as expected, and the trailing tail follows the precession of the satellite's orbit.\\
(vi) As Johnston et al. (2005) have shown, the precession of the orbital plane is highly sensitive to the halo's shape. Defining $\Delta\phi\equiv\langle \phi\rangle_l-\langle \phi\rangle_t$, where $\langle\phi\rangle_l, \langle\phi\rangle_t$ are the azimuthal angles of the leading and trailing tails, we find that ${\rm d}\Delta \phi/{\rm d}t\simeq 40^\circ, 25^\circ, 12^\circ$ per Gyr for $q=0.8, 1.0 $ and $ 1.2$, respectively (in contrast, the progenitor's precession rate was $30^\circ, 14^\circ, -2^\circ$ per Gyr).

\subsection{The satellite mass and the stream's width and velocity dispersion}
As shown by Johnston (1998), the width and the velocity dispersion perpendicular to the orbital plane of the youngest stream parts provide strong constraints on the mass of the progenitor. As the age increases, streams become wider and ``hotter'' (e.g. Helmi \& White 1999), with properties reflecting the shape of the galaxy potential.

\begin{figure}
\plotone{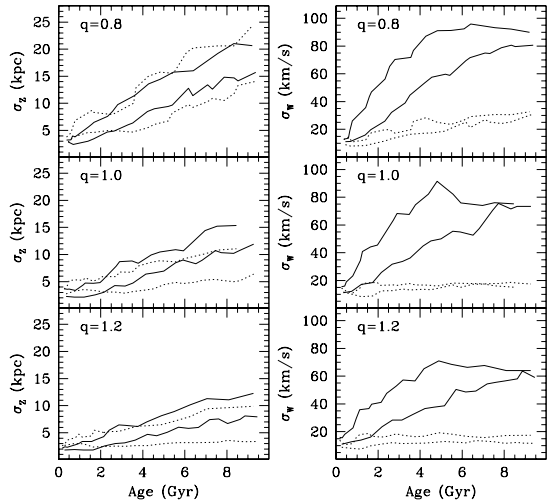}
\caption{Width of the tidal tails (left columns) and their velocity dispersion (right column) at $t=t_f$, measured perpendicular to the orbital plane as a function of age. Solid and dotted lines denote particles in the leading and trailing tails, respectively. Models M1 and M2 are shown with thin and thick lines, respectively. These models assume a static halo.}
\label{fig:zw}
\end{figure}

In Fig.~\ref{fig:zw} we plot the width ($\sigma_Z$) and the velocity dispersion ($\sigma_W$) in the direction perpendicular to the stream's orbital plane (denoted here as the $Z$-axis) as a function of age. These quantities are defined simply as $\sigma_x\equiv\langle (x-\langle x\rangle)^2\rangle^{1/2}$.
We divide the sample of stream particles at present into leading and trailing tails (solid and dotted lines, respectively) and into sub-samples of different ages. For each sub-sample, we calculate the mean orbital plane in order to determine its perpendicular vector.
We use thin and thick lines to represent values from the models M1 and M2. This Figure shows that:\\
(i) Young tidal streams (age $< 1$ Gyr) present similar width and velocity dispersion values, $\sigma_Z\simeq 2$--$3$ kpc and $\sigma_W\simeq 10$--$15$ km/s, barely dependent on the progenitor's mass evolution and on the halo's axis-ratio, in agreement with Johnston (1998).\\
(ii) Older stream parts show a clear increase of width and velocity dispersion as a function of age. That increase is larger in streams originating from massive progenitor systems and from  progenitors orbiting in an oblate halo. Distinguishing between leading and trailing tails we obtain that the former are wider than the later for age $> 2$ Gyr. Curiously, all models show that for $1 < {\rm age} < 2$ Gyr the trailing arm is slightly wider than the leading one.\\
(iii) The stream's velocity dispersion has a complex evolution. Whereas the trailing tail shows ${\rm d}\sigma_W/{\rm d}t\simeq {\rm const}>0$ for the whole age range, the leading tail has a steep rise of $\sigma_W$ followed by a plateau at age$\sim 5$--$6$ Gyr. The value of $\sigma_W$ at the plateau is comparable to the velocity dispersion of halo stars (see e.g. Binney \& Merrifield 1998), which indicates that the kinematics of stars that belong to the oldest stream pieces are dominated by the Galaxy potential, regardless of their ``external'' origin.  

As Ibata et al. (2001) showed, the thickness and velocity dispersion are directly correlated with the precession rate of the tidal stream. By comparing Fig.~\ref{fig:phi} and~\ref{fig:zw} we can readily see that larger precession rates always induce tidal streams to become thicker and ``hotter''. 

In the absence of proper motions, tidal streams are usually identified from the background stellar population as spatial over-densities with a ``cold'' (i.e small $\sigma_W$) velocity distribution. Our results appear to indicate that, whereas the trailing tails can be easily detected, the old parts of the leading tails might be extremely hard to distinguish from halo stars if the only constraints available are the two mentioned above.

\subsection{Streams in the $E-L_z$ plane. Is secular halo evolution detectable?}\label{sec:el}
Full observational coverage of a tidal stream is a difficult task, firstly because tidal streams are spread out over a large range of distances and angles (therefore requiring large observational surveys, such as SDSS, RAVE or GAIA) and secondly because it is difficult to distinguish between tidal stream and Galactic stars, especially for old parts of the stream. As Helmi \& White (1999) showed, tidal streams populate well-defined regions of the constant-of-motion space, independent of their spatial distribution and age, which can enormously facilitate their identification.

\begin{figure}
\plotone{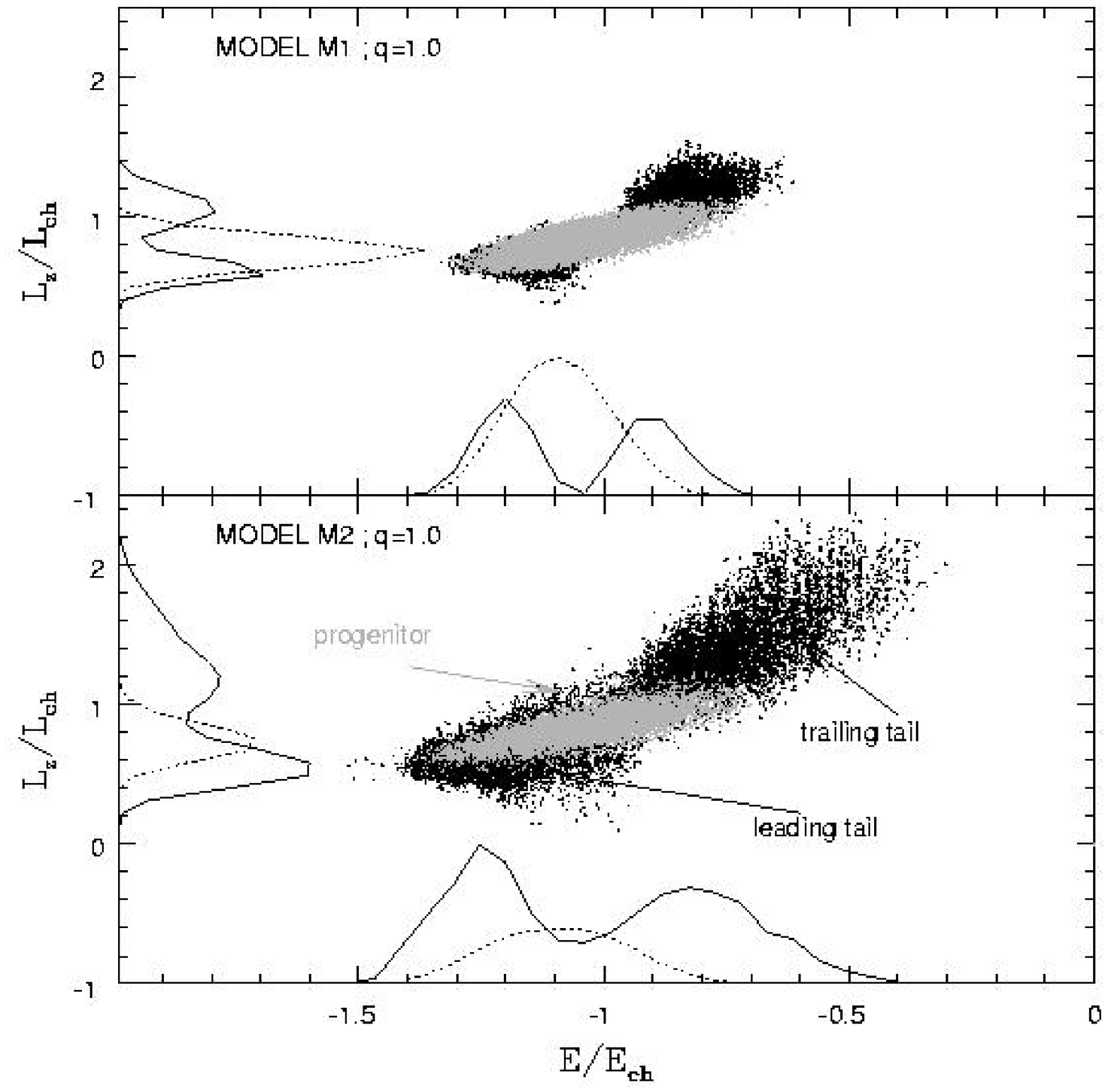}
\caption{$E-L_z$ distribution of the particles in the satellite and the two tidal tails at $t=t_f$, for model M1H1Q2 (upper panel) and M2H1Q2 (bottom panel). Solid and dotted lines show the distribution of tidal stream and bound particles, respectively. All distributions have been normalized to the number of satellite particles ($N=10^5$).}
\label{fig:led}
\end{figure}

In Fig.~\ref{fig:led} we show the present location in the energy-angular momentum plane of the N-body models M1H1Q2 (upper panel) and M2H1Q2 (bottom panel) at $t=t_f$. Both models were evolved in a spherical, static halo. Grey dots show particles that remain bound at the end of the simulation, whereas black dots denote tidal stream particles. Tidal streams show a bimodal $E$ and $L_z$ distribution, corresponding to the leading and trailing tails. In comparison with the remnant system, leading tails have lower $E, L_z$ mean values and trailing tails higher values.

Satellite galaxies that were initially more massive have suffered a larger orbital decay through dynamical friction and show, therefore, a larger spread in the constant-of-motion plane (compare model M2 against model M1).
Remarkably, the area in the $E-L_z$ plane occupied by bound particles only depends on the present mass and not on the mass loss history. That fact demonstrates that the mass evolution of a bound stellar system can be hardly estimated exclusively from the present $E$--$L_z$ distribution of bound particles.

\begin{figure}
\plotone{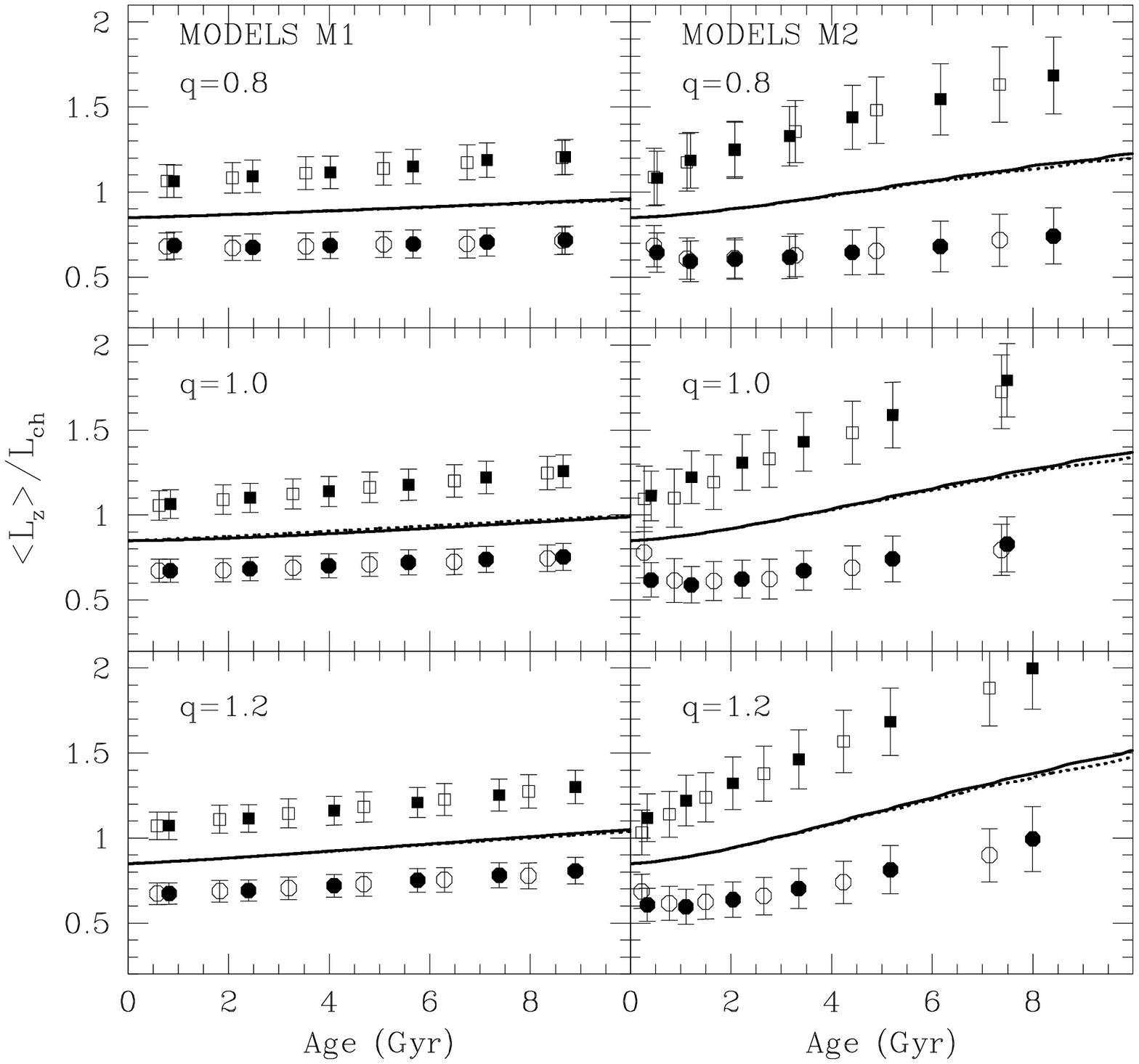}
\caption{Averaged angular momentum at $t=t_f$ as a function of stream age. Error bars represent the variance around the mean value. Open and filled symbols denote stream particles that have orbited in static and evolving halos, respectively. Squares (circles) represent trailing (leading) particles. Solid (dotted) thick lines show the angular momentum of the progenitor's orbit as a function of look-back time for simulations in static (evolving) halos. Note that particles in evolving and static halos show the same angular momentum distribution because $L_z$ is an adiabatic invariant.}
 
\label{fig:lm1m2}
\end{figure}

\begin{figure}
\plotone{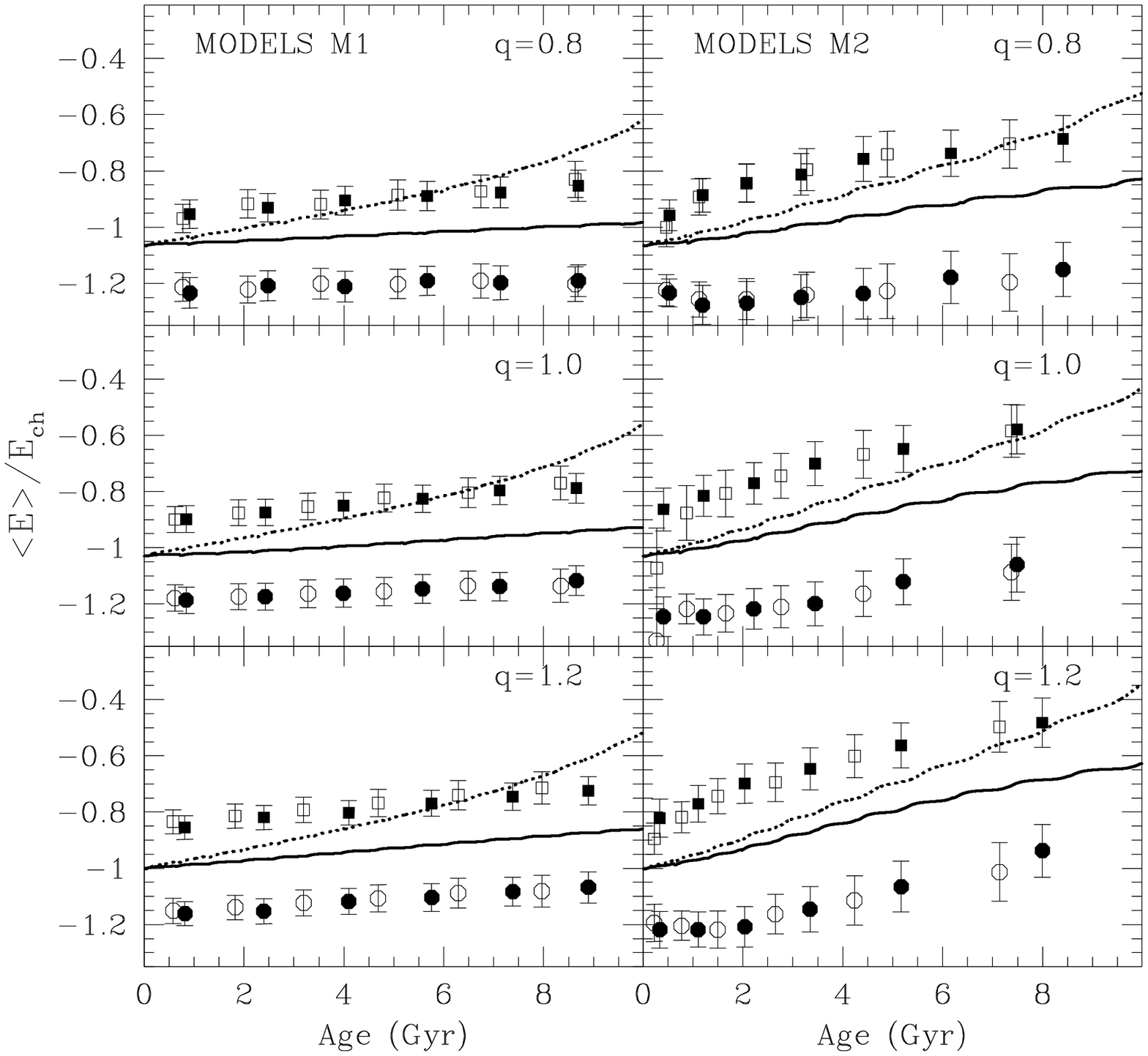}
\caption{Average energy at $t=t_f$ as a function of stream age. We use the notation of Fig.~\ref{fig:lm1m2}. Note that particles in evolving (filled symbols) and static (open symbols) halos show at present the same energy distribution, even though the satellite orbital energy was different in the past (compare dotted against solid lines, respectively).}
\label{fig:em1m2}
\end{figure}

We have separated the present distribution of tidal stream particles into bins of different ages and measured the mean energy and angular momentum of the leading and trailing tails. In Fig.~\ref{fig:lm1m2} and~\ref{fig:em1m2} we plot those quantities for all our satellite models. Thick lines also show the look-back evolution of the progenitor's energy and angular momentum (solid and dotted lines denote models in static and evolving halos, respectively). In this Figure we can see that:\\
(i) The evolution of the progenitor's energy and angular momentum induced by dynamical friction can be easily traced from tidal streams if we know the time when stream particles were stripped out. This ultimately allows us to determine accurately the mass history of satellite galaxies with associated tidal streams.\\
(ii) The present energy and angular momentum of stream particles are not sensitive to the adiabatic growth of the dark matter halo, independent of the time when stream particles were stripped from the progenitor. \\
As shown in Fig.~\ref{fig:lm1m2}, the angular momentum of the main system is independent of whether the halo evolves or not, since $L_z$ is an adiabatic invariant, which is also applicable to stream (unbound) particles. In contrast, Fig.~\ref{fig:em1m2} shows that the energy of the main system followed a fairly different evolution in evolving and static halos (compare thick dotted and solid lines). However, that is not reflected in the {\it present} energy distribution of the stream, independent of its age (compare filled and open symbols).

\begin{figure}
\plotone{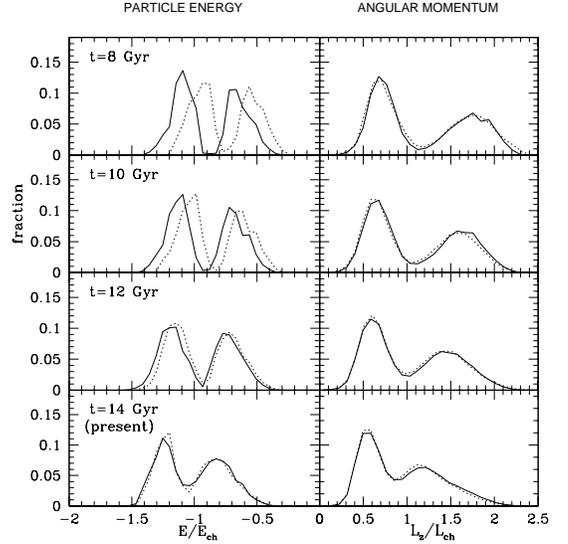}
\caption{Energy (left) and angular momentum (right) distribution of the simulations M2H1Q2 (solid lines; static halo) and M2H2Q2 (dotted lines; evolving halo) at different snapshots. All curves are normalized to the number of stream particles at the given time.}
\label{fig:elt}
\end{figure}

The reason is that the energy of the stream particles re-adjust at any time-step to the evolution of the halo's potential. This can be clearly seen in Fig.~\ref{fig:elt}, where we plot the $E, L_z$ distributions at $t=8,10,12$ and 14 Gyr. Solid and dotted lines show the distributions from the models M2H1Q2 (static halo) and M2H2Q2 (evolving halo).
As we commented above, $L_z$ is insensitive to the time-dependence of the Galaxy potential. In contrast, the energy distribution of stream particles varies as a result of the halo's growth.\\
This figure shows that, at any given time, the mean value of the stream energy is approximately that of the progenitor system (see thick lines in Fig.~\ref{fig:em1m2}). 
As a result, the present energy of stream particles is determined exclusively by the present Galaxy potential and do not depend on the way the Galaxy has evolved.
In other words, tracking back the halo evolution using tidal streams is a degenerate problem, as we have only information at the present day.

\subsection{Spatial distribution of debris}\label{sec:xyz}
\begin{figure}
\plotone{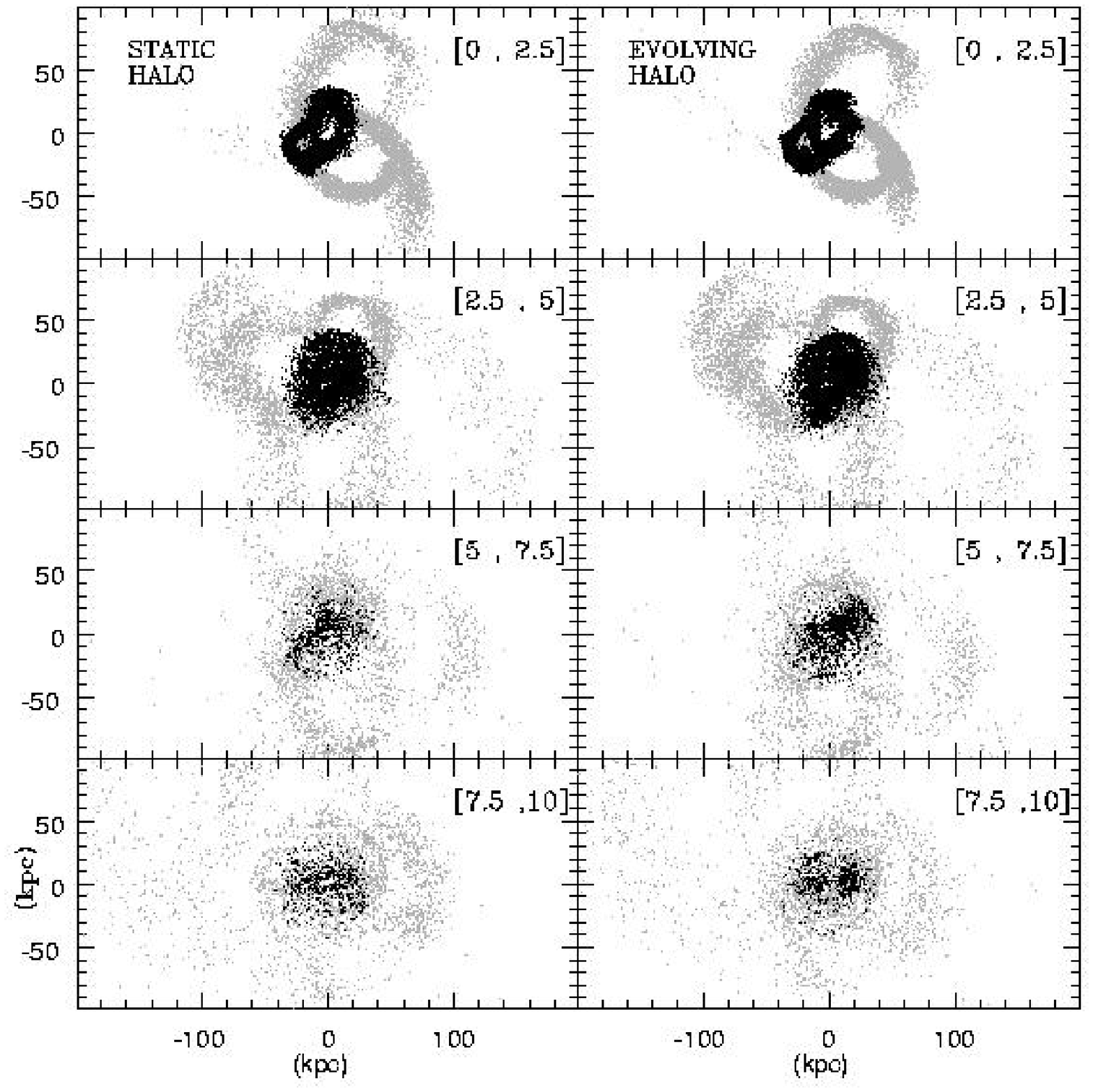}
\caption{Projection into the present orbital plane of the present distribution of tidal stream particles for the models M2H1Q2 (static halo with $q=1$, left column) and M2H2Q2 (evolving halo with $q=1$, right column).
Grey and black dots represent trailing  and leading tail particles, respectively. We have divided the particle sample into different age intervals (see upper-right corners, the age is given in gigayears).}
\label{fig:xyz}
\end{figure}

Two tidal streams with the same $E$--$L_z$ distribution and with their progenitor systems at the same position will present the same spatial and kinematical properties. In order to show that, we plot in Fig.~\ref{fig:xyz} the spatial projection into the orbital plane of the present distribution of debris for the models M2H1Q2 (static halo, $q=1$, left panels) and the model  M2H2Q2 (evolving halo, $q=1$, right panels). We have divided each model into leading (grey dots) and trailing (black dots) particle samples. Additionally, each row shows the particle distribution for different stream age intervals (given in the upper-right corner in gigayears).\\
This figure shows interesting features:\\
(i) Leading tail particles move, on average, at closer Galactocentric distances than trailing tail particles, as one would expect from their lower energy and angular momentum.\\
(ii) Trailing tails keep their coherent structure even if they were stripped 10 Gyr ago, whereas leading tails disperse within 2.5 Gyr, making their detection as spatial over-densities more difficult. If we compare these results with those shown in Fig~\ref{fig:phi}, we come to the conclusion that the ``broadening'' and ``heating'' of tidal streams is more prominent in systems that suffer a large precession.
(iii) As one expects from the results of \S\ref{sec:el}, the present spatial distributions of debris from progenitors that have orbited in static and evolving halos are essentially indistinguishable if we force the progenitor to end up with the same mass, position and velocity for all our models.

\section{Effects of the halo shape evolution}\label{sec:shape}
In this contribution we have analyzed the effects that the growth in mass and size of the Galaxy induce on the stream's properties. However, we have assumed that the shape of the halo remains constant in time. 
According to recent cosmological simulations, that assumption appears to be false. Dissipationless $\Lambda$CDM simulations  show that, as a result of accretion, dark matter halos become more spherical in time (see e.g. Allgood et al. 2005 and references therein). For a Milky Way-like galaxy, these authors show that $\langle q\rangle(z=3)\simeq 0.4$, increases up to $\langle q\rangle(z=0)\simeq 0.6$, whereas $d\langle q\rangle/dr\simeq 0$. Dissipation enhances that process (Kazantzidis et al. 2004, Springel et al. 2004, Bailin 2005), specially in the innermost region of the galaxy ($r<0.3 r_{\rm rvir}$), although hydrodynamical simulations appear to suffer from over-cooling, which over-estimates dissipation effects.

\begin{figure}
\plotone{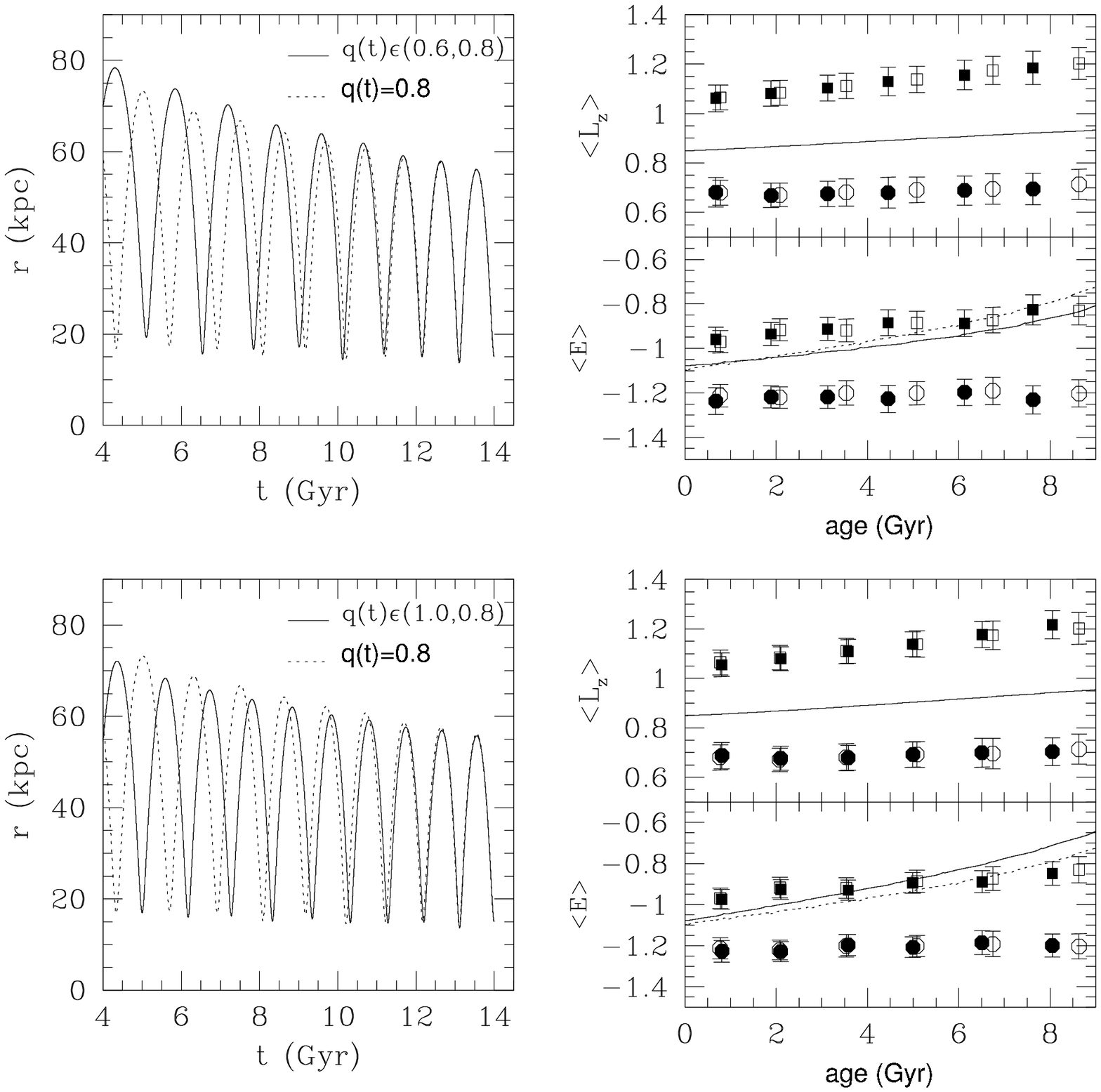}
\caption{\emph{Left column}: Radius evolution. \emph{Top panel}: Averaged $L_z$ normalized to $L_{\rm ch}$ obtained from the present distribution of debris as a function of stream age.  \emph{Upper-middle panel}: Averaged energy normalized to $E_{\rm ch}$ as a function of stream age. In these panels open dots denote simulations in a halo with $q=0.8={\rm const.}$, whereas filled dots denote debris in a halo with axis-ratio that increases linearly in time from $q=0.6$ up to $q=0.8$.   In the bottom and middle-bottom panels  we show the same quantities for satellites evolving in a  halo with axis-ratio that decreases linearly in time from $q=1.0$ down to $q=0.8$ (filled dots) and a halo with $q=0.8={\rm const.}$ (open dots). In all panels, full and dotted lines represent the progenitor's quantities for a varying and a constant $q$, respectively. The initial and final mass of the satellite model are $M_s(t_0)=10^9 M_\odot$ and $M_s(t_f)=5\times 10^8 M_\odot$, respectively. Note that all simulations use an evolving halo. }
\label{fig:rel_q}
\end{figure}

In order to address whether the halo shape evolution can be uncovered from tidal streams, we have repeated our simulations in halos with a variable axis-ratio. For simplicity, we have assumed that $q$ varies linearly with time  from $q(t_0)=0.6$ up to $q(t_f)=0.8$ (i.e., the halo becomes more spherical with time) and from $q(t_0)=1.0$ down to $q(t_f)=0.8$ (i.e, the halo becomes oblate with time). 

In Fig.~\ref{fig:rel_q} we show the radius evolution (left column) for both cases. Additionally, we plot the evolution of the model M1H2Q1 (i.e with $q=0.8=$const.) for comparison. As expected from Fig.~\ref{fig:rapoe}, increasing the value of $q$ induces a reduction of the satellite's peri and apo-galactica, and {\it vice versa}. \\
In the right column we plot the present distributions of energy and the angular momentum of tidal debris as a function of stream age. Filled and open dots denote debris in a halo with varying and a constant axis-ratio, respectively, whereas the error bars represent the variance around the mean value. Full and dotted lines show the look-back evolution of the progenitor's energy and angular momentum.\\
The results that we obtain are clearly equivalent to those of \S\ref{sec:el}: Although the orbit of the stream progenitor depends on the evolution of the halo potential, the present distribution of debris only reflects the {\it present} Galaxy potential . In particular, Fig.~\ref{fig:rel_q} shows the energy and angular momentum distributions are insensitive to the evolution of the halo's shape if we force our models to end up with the same mass, position and velocity.  

\section{Effects of dark matter clumps on tidal stream properties}\label{sec:clumps}
Having explored the effects of secular evolution of the smooth host potential, we now turn to the role of dark matter clumps in dispersing unbound (stream) particles. For simplicity, we assume that the progenitor's orbit remains exclusively determined by the smooth component of the Galaxy potential, so that $f_{\rm sub}$ (Eq.~\ref{eq:f_sub}) only applies to unbound satellite particles. 

Note that in self-consistent simulations, the satellite's orbit can also be altered through strong satellite--subhalo interactions, which would be  easily detectable in the present stream's energy-angular momentum distribution as discontinuities at different stream ages\footnote{We refer the reader to Pe\~narrubia \& Benson (2005) for a statistical analysis of the effects and likelihood of clump--clump interactions in a Milky Way-like galaxy. These authors find that the present population of subhalos move on orbits that, statistically, have been barely altered by subhalo-subhalo interactions.}. Therefore, our approach minimizes the effect of dark matter clumps and only analyzes the heating induced by the halo's lumpiness on tidal streams.

We have carried out simulations with nine different sets of subhalos (see \S\ref{sec:code}) drawing on the method in \S\ref{sec:mtree} to construct merger trees and assuming an isotropic accretion of subhalos. Thus, for consistency, we consider only a spherical dark matter halo which evolves according to Eq.~(\ref{eq:mhevol}), (\ref{eq:c})  and~(\ref{eq:rv}). 

For the parent satellites, we consider the orbits of the models M1H2Q2 and M2H2Q2. For each merger tree, we evolve these orbits and analyze the present-day $E$ and $L_z$ distributions. Since the time required for the subhalo force calculation scales as $N_s^2$, where $N_s$ is the total number of subhalos, and since in $\Lambda$CDM the number of subhalos in a given mass range goes as $n_s(M)dM\sim M^2 dM$, our study is limited to subhalos with masses above $10^7 M_\odot$ (which implies a typical $\log_{10} N_s\sim 3$) .

\begin{figure}
\plotone{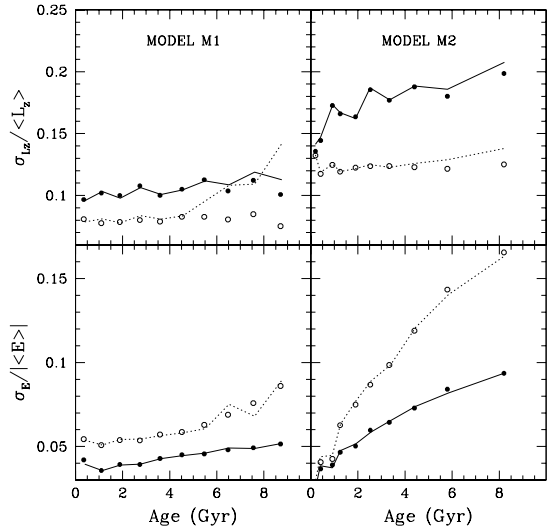}
\caption{Variance of the present energy (bottom panels) and angular momentum (upper panels) distributions as a function of stream age. Filled and open dots denote, respectively, the variance of the leading and trailing tails in the absence of dark matter clumps. Solid and dotted lines show, respectively, the variance of the leading and trailing tails in galaxies with a subhalo population corresponding to that predicted by $\Lambda$CDM cosmology for a Milky Way-like galaxy.}
\label{fig:el_clump}
\end{figure}

By construction, the only effect of repeated encounters between tidal stream particles and subhalos is a  progressive broadening of the stream distribution in the $E,L_z$ plane.  In Fig.~\ref{fig:el_clump} we have plotted the ratio between the variance and the averaged value of the energy and the angular momentum as a function of the stream age for the models M1H2Q2 (left column) and M2H2Q2 (right column) in the absence (dots) and presence (lines) of dark matter clumps. We find that:\\
(i) As a result of encounters between subhalos and stream particles, the angular momentum dispersion shows a clear increase in the trailing tail (upper-left panel, dotted line) for age$>5$ Gyr, whereas the angular momentum distribution of the leading tail (full line) has a dispersion that remains practically unaltered. That increase can be as large as a factor 1.75 for the model M1, reducing to a factor 1.07 for the model M2. Remarkably, both models show that the stream heating can only be observed for those parts older than 5 Gyr (i.e., $t<9$ Gyr). \\
(ii) We do not observe any increase of $\sigma_E$ induced by subhalos. This results is in agreement with that found by Ibata et al. (2002), who showed that stream heating resulting from subhalo interactions is mostly reflected in the angular momentum distribution, whereas the energy distribution is barely altered.

The angular momentum dispersion only shows a remarkable increase for age$>$5 Gyr.
Following the results of Pe\~narrubia \& Benson (2005)-- see also Zentner \& Bullock (2003)-- the accretion rate of substructures found a maximum at $t=4$ Gyr, decreasing a factor 3 until $t=9$ Gyr. From $t=9$ Gyr to the present, the accretion rate was approximately constant. \\
This appears to indicate that the stream heating is correlated with the accretion rate of subhalos:  We observe heating only for age$>5$ Gyr because the encounter rate between stream particles and dark matter clumps was approximately a factor 3 larger for those particles stripped at $t<9$ Gyr (age$>$5 Gyr) than for those that became unbound at $t>9$ Gyr (age$<$5 Gyr).

Finally, we must remark that the halo's lumpiness will be more easily detected in the old (age$>5$ Gyr) and cold (trailing tails of progenitors that were initially low mass) parts of tidal streams.

\section{Discussion and conclusions}\label{sec:disc}
In this paper we have analyzed what information can be extracted from stellar tidal streams on (i) the Milky Way's halo shape, (ii) the halo's secular evolution, (iii) the mass evolution of the stream's progenitor and (iv) the presence of dark matter clumps in our Galaxy.
We assumed as a boundary condition of this analysis that the present-day position, velocity and mass of the stream's progenitor can be measured and are identical for all evolutionary scenarios. Under these conditions, we have explored whether the extended tidal debris reflect differences arising from the items (i)--(iv) above.
We have carried out this study for a Sagittarius-like dwarf galaxy, although our results are general and can be applied to other systems in the Milky Way.

The main result is that tidal streams do not provide information on the adiabatic evolution of the Milky Way or, in other words, the properties of entire tidal streams only reflect the {\it present} Galaxy potential. 
Thus, ground-based observations already available for tidal streams (basically providing distances and radial velocities along the stream) and future satellite data covering the full phase-space (making possible studies in the $E$--$L_z$ plane) can only constrain the present characteristics of the Milky Way potential. As a direct consequence, Galaxy evolution processes can be neglected when modeling tidal streams, which clarifies one of the main caveats in current N-body simulations and confirms that tidal stream models computed under that hypothesis are appropriate  for tracing the distribution of dark matter around our Galaxy.

In contrast, we find that tidal streams are indeed sensitive to the present properties of halo shape. In particular, we confirm that measuring the precession rate is a fairly powerful method to constrain the halo flattening of the gravitational potential. This may not require measurements of proper motions.

We have shown that the study of tidal streams in the $E$--$L_z$ plane provides information on the progenitor's mass loss since the time of accretion.  The energy--angular momentum distribution of stream particles has an average value that only reflects the present position and velocity of the progenitor system. In contrast, the $E$--$L_z$ variance about that mean increases with the initial satellite mass, thus, making it possible to determine the mass loss fraction directly from the present $E$--$L_z$ distribution. Furthermore, since a secular drift in energy and angular momentum is induced by dynamical friction (a drag acceleration that scales in proportion to $M_{\rm s}$) one could, in principle, reconstruct the mass loss curve $M_s(t)$ from the age of different stream pieces (if the age can be labeled e.g. by metalicity or by theoretical modeling), which ultimately depends on the initial mass profile of the satellite galaxy (see Zhao 2004).

We have analyzed the effects of dark matter clumps on tidal streams in Section~\ref{sec:clumps}. We simplify the problem by assuming that dark matter clumps do not alter the progenitor's orbit, but only the orbits of stream particles. That approach establishes, therefore, a minimum impact of subhalos on tidal stream properties (we note that a sharp change of the progenitor's $E_p, L_{z,p}$ induced by a collision with a subhalo at $t=t_c$ would be reflected as a discontinuity in the averaged stream's $E, L_z$ at age=$t_f-t_c$).
 We have confirmed that dark matter subhalos induce only very modest stream ``heating'' by increasing of the angular momentum dispersion in the oldest (age$>$5 Gyr) and coldest (trailing tail) stream parts.\\
This raises the question of whether one can constrain the halo lumpiness either with current ground-based techniques or with future astrometric satellite missions (GAIA, SIM).\\
At present, the detection of the oldest parts (i.e those stars that became unbound first) of a tidal stream with state-of-the-art ground-based surveys is challenging. Although the trailing tail maintains a coherent structure and
should be easier to detect as spatial over-densities (see Sec.~\ref{sec:xyz}), its surface brightness decreases considerably with time, which reduces the possibility of
 detection above the Galactic field contamination  in large field-of-view color-magnitude diagrams like those provided by SDSS. Furthermore, it is also expected that tidal streams are composed by old, metal-poor stellar population. 
Therefore, some  valuable techniques for tracing tidal streams with all-sky surveys (like 2MASS) cannot be applied to detect the oldest stream pieces since these are barely sensitive to metal-poor stars expected in old parts of tidal streams. Also surveys of tidal streams using M-giant stars (Majewski et al. 2004) are limited to the youngest stream pieces (material unbound only 1-2 Gyr ago; LJM), while the oldest wraps still remain hidden in
the Galactic halo. Current theoretical models of the two largest, brightest streams in the Milky Way (Sgr: LJM;  Monoceros: Pe\~narrubia et al. 2005) indicate that there is no detection of tidal debris that became unbound more than 2-3 Gyr ago.  Finally, stream tails are not located on the progenitor's orbital plane (see Sec.~\ref{sec:orbplane}), which further complicates tagging of debris as part
of a known stellar system and needs of accurate stream models in order to determine a possible common origin and to estimate the stream's age. \\
In the future, the most powerful method to search for ancient debris in the halo should come with the next generation of astrometric satellites, which will permit analysis of tidal streams in the constant-of-motion space, thus, providing the most straightforward way to identify tidal debris independently of the stream age. However, it is important to remark that observational errors may introduce strong limitations, as indicated by Brown et al. (2005). Possibilities of identifying satellite remnants in the Milky Way halo are considerably reduced after taking into account the observational errors expected for the GAIA catalogue and the large number of background stars.

\section*{Acknowledgements}
We would like to thank R.Ibata for his helpul comments on this paper and S.Kazantzidis for his insights on the evolution of DMH shapes. 
AJB acknowledges support from a Royal Society URF.

{}

\begin{thebibliography}{}
\bibitem[]{}Allgood B., Flores R.A., Primack J.R., Kravtsov A.V., Wechsler R., Faltenbacher A., Bullock J.S., 2005, astro-ph/0508497
\bibitem[]{}Bailin J., Kawata D., Gibson B., Steinmetz M., Navarro J., Brook C., Gill S., Ibata R., Knebe A., Lewis G., Okamoto T., 2005, ApJL, 627, L17
\bibitem[]{}Barden et al., 2005, astro-ph/0502416
\bibitem[]{}Benson A.J., Lacey~C.G., Baugh C.M., Cole S., Frenk C.S., 2002, MNRAS, 333, 156
\bibitem[]{}Benson A.J., Lacey C.G., Frenk C.S., Baugh C.M., Cole S., 2004, MNRAS, 351, 1215
\bibitem[]{}Binney J., 1977, MNRAS, 181, 735
\bibitem[]{}Binney J., Tremaine S., 1987, Galactic Dynamics. Princeton University Press, Princeton, New Jersey 
\bibitem[]{}Binney J., Merrifield M., 1998, Galactic Astronomy. Princeton University Press, Princeto
n, New Jersey
\bibitem[]{}Brown A.G.A., Vel\'azquez H.M., Aguilar L.A., 2005, 359, 1287
\bibitem[]{}Bryan G., Norman M., 1998, ApJ, 495, 80
\bibitem[]{}Bullock J., Kravtsov A., Weinberg D., 2000, ApJ, 539, 517
\bibitem[]{}Chandrasekhar S., 1960, Principles of Stellar Dynamics. Dover, New York
\bibitem[]{}Dav\'e R., Spergel D.N., Steinhardt P.J., Wandelt B.D., 2001, ApJ, 547, 574
\bibitem[]{}Dubinsky J., Calberg R.G., 1991, ApJ, 378, 496
\bibitem[]{}Dubinsky J., 1994, ApJ, 431, 617
\bibitem[]{}Fellhauer M., Kroupa P., Baumgardt H., Bien R., Boily C. M., Spurzem R., Wassmer N., 2000, NewA, 5, 305
\bibitem[]{}Font A.S., Johnston K.V., Guhathakurta P., Majewski S.R., Rich R.M., 2004, astro-ph/0406146
\bibitem[]{}Hayashi E., Navarro J., Taylor J.E., Stadel J., Quinn T., 2003, ApJ, 584, 541
\bibitem[]{}Helmi A. \& White S.D.M., 1999, MNRAS, 307, 495 
\bibitem[]{}Hernquist L., 1990, ApJ, 356, 359 
\bibitem[]{}Ibata R., Irwin M., Lewis G.F., Stolte A., 2001, ApJ, 547, 133
\bibitem[]{}Ibata R.A., Lewis G.F., Irwin M.J., Quinn T., 2002, MNRAS, 332, 915
\bibitem[]{}Irwin M., Hatzidimitriou D., 1995, MNRAS, 277, 1354
\bibitem[]{}Johnston K.V., 1998, ApJ, 495, 297
\bibitem[]{}Johnston K.V., Majewski S.R., Siegel M.H., Reid I.N., Kunkel W.E., 1999, AJ, 118, 1719
\bibitem[]{}Johnston K.V., Sackett P.D., Bullock J.S., 2001, ApJ, 557, 137
\bibitem[]{}Johnston K.V., Spergel D.N. Haydn C., 2002, ApJ, 570, 656
\bibitem[]{}Johnston K.V., Law D.R., Majewski S., 2005, ApJ, 619, 800
\bibitem[]{}Just A., Pe\~narrubia J., 2005, A\&A, 431, 861
\bibitem[]{}Kazantzidis S., Kravtsov  A., Zentner A.R., Allgood B., 2004, ApJL, 611, L73
\bibitem[]{}King I.R., 1966, AJ, 71, 65
\bibitem[]{}Kleyna J., Wilkinson M.I., Evans N.W., Gilmore G., Frayn C., 2005, MNRAS, 330, 792
\bibitem[]{}Klypin A., Kravtsov A., Valenzuela O., Prada F., 1999, ApJ, 522, 82
\bibitem[]{}Klypin A., Zhao H., Somerville R., 2002, ApJ, 573, 597
\bibitem[]{}Law D.R., Johnston K.V., Majewski S.R., 2004, astro-ph/0407566 (LJM)
\bibitem[]{}Lokas E. W., Mamon G.A., Prada F., 2005, MNRAS, 363, 918
\bibitem[]{}Lynden-Bell D. \& Lynden-Bell R.M., 1995, MNRAS, 275, 429
\bibitem[]{}Mateo M.L., 1998, ARA\&A, 36, 435
\bibitem[]{}{\bf Miyamoto ref!!!}
\bibitem[]{}Moore B., Ghigna S., Governato F., Lake G., Quinn T., Stadel J., Tozzi P., 1999, ApJ, 524, L19
\bibitem[]{}Mu\~noz R., Frinchaboy P. M., Majewski S. R., Kuhn J. R., Chou M., Palma C., Sohn S. T., Patterson R. J., Siegel M. H., 2005, ApJ, 631L, 137
\bibitem[]{}Navarro J., Frenk C.S., White S.D.M., 1995, MNRAS, 275,56
\bibitem[]{}Navarro J., Frenk C.S., White S.D.M., 1996, ApJ, 462, 563
\bibitem[]{}Navarro J., Frenk C.S., White S.D.M., 1997, ApJ, 490, 493
\bibitem[]{}Pe\~narrubia J., Mart\'inez-Delgado D., Rix H.W., G\'omez-Flechoso M.A., Munn J., Newberg H., Bell E., Yanny B., Zucker D., Grebel E., 2005, ApJ, 626, 128
\bibitem[]{}Pe\~{n}arrubia J., Just A., Kroupa P., 2004, MNRAS, 349, 747
\bibitem[]{}Pe\~narrubia J., Benson A.J., 2005, MNRAS, 364, 977
\bibitem[]{}Pohlen M., Mart\'inez-Delgado D., Majewski S., Palma C., Prada F., Balcells M., 2004, ASPC, 327, 288
\bibitem[]{}Sackett P.D., 1999, in Merritt D., Sellwood J.A., Valluri M., eds, ASP Conf. Ser 182, Galaxy Dynamics, Astron. Soc. Pac., San Francisco, p.393
\bibitem[]{}Somerville R.S., 2002, ApJ, 572, 23
\bibitem[]{}Springel V., White S.D.M., Hernquist L., 2004, in IAU symposium The shapes of simulated dark matter halos, p.241
\bibitem[]{}Taylor J.E., Babul A., 2001, ApJ, 559, 716
\bibitem[]{}Trujillo I., Pohlen M., 2005, ApJ, 630, 17
\bibitem[]{}Tully B., Somerville R., Trentham N., Verheijen M., 2002, ApJ, 569, 573
\bibitem[]{}Wechsler R.H., Bullock J.S., Primack J.R., Kravtsov A.V., Dekel A., 2002, ApJ, 568, 52
\bibitem[]{}Yoshida N., Springel V., White S.D.M., Tormen G., 2000, ApJ, 535, L103
\bibitem[]{}Zhao H., 2004, MNRAS, 351, 891      
\bibitem[]{}Zentner A.R., Bullock J., 2003, ApJ, 598, 49
\end{thebibliography}
\end{document}